\documentclass[10pt]{article}
\usepackage{mathrsfs}
\usepackage{natbib,graphicx,setspace,lscape,longtable}
\usepackage{mathrsfs,amsmath,amsthm,amssymb,color}
\usepackage{natbib,epsfig,graphicx,pdfpages}
\usepackage{hyperref}
\usepackage{bm}
\usepackage{epsfig,graphicx}
\usepackage{caption,subcaption}
\bibpunct{(}{)}{;}{a}{,}{,}

\newtheorem{theorem}{Theorem}
\newtheorem{lemma}{Lemma}
\newtheorem{proposition}{Proposition}
\def\beq{\begin{equation}}
\def\eeq{\end{equation}}
\def\beqr{\begin{eqnarray}}
\def\eeqr{\end{eqnarray}}
\def\beqrs{\begin{eqnarray*}}
\def\eeqrs{\end{eqnarray*}}
\def\bet{\begin{theorem}}
\def\eet{\end{theorem}}
\def\bel{\begin{lemma}}
\def\eel{\end{lemma}}
\def\bep{\begin{proposition}}
\def\eep{\end{proposition}}
\def\bg{\begin{figure}[tbph]\begin{center}}
\def\eg{\end{center}\end{figure}}

\def\bc{\begin{center}}
\def\ec{\end{center}}

\def\bX{\mathbf{\mathbf{X}}}

\def\mD{\mathcal D}

\def\mR{\mathbb{R}}
\def\mS{\mathcal S}

\def\mZ{\mathbb{Z}}

\def\brho{\boldsymbol{\rho}}

\def\cov{\mbox{cov}}

\def\bSigma{\bm{\Sigma}}

\def\brho{\bm{\rho}}

\def\bX{\mathbf{X}}

\usepackage{geometry}
\numberwithin{equation}{section}

\def\spacing{1}
\begin{document}

\title
{\Large Multiple Influential Point Detection in High-Dimensional Spaces}
\author{Junlong Zhao\quad\quad
 Chao Liu \quad\quad
 Lu Niu \quad\quad
  Chenlei Leng\footnote{Zhao is Associate Professor, School of Statistics, Beijing Normal University, China. Liu and Niu are graduate students, School of Systems and Mathematics, Beihang University, China. Leng is Professor, Department of Statistics, University of Warwick, UK. Leng is also affiliated with the Alan Turing Institute.
 Corresponding author: Chenlei Leng (Email:  C.Leng@warwick.ac.uk).} }

\date{\today}
\maketitle
\def\baselinestretch{1}\selectfont
 \begin{abstract}
Influence diagnosis is an integrated component of data analysis, but is severely under-investigated in a high-dimensional setting. One of the key challenges, even in a fixed-dimensional setting, is how to deal with multiple influential points giving rise to the masking and swamping effects. This paper proposes a novel group deletion procedure referred to as MIP by studying two extreme statistics based on a marginal correlation based influence measure. Named the Min and Max statistics, they have complimentary properties in that the Max statistic is effective for overcoming the masking effect while the Min statistic is useful for overcoming the swamping effect. Combining their strengths, we further propose  an efficient algorithm that can detect influential points with a prespecified false discovery rate.      The proposed influential point detection procedure is simple to implement, efficient to run, and enjoys attractive theoretical properties. Its  effectiveness is verified empirically via extensive simulation study and data analysis. An R package implementing the procedure is freely available.

\end{abstract}
\noindent {\bf Keywords}: False discovery rate, group deletion, high-dimensional linear regression, influential point detection, masking and swamping, robust statistics. 

\noindent{\bf Running Title}: Multiple Influential Point Detection.

%\newpage

\def\baselinestretch{\spacing}\selectfont

\section{Introduction}
The last few decades have witnessed an explosion of high-dimensional data in applied fields including biology, engineering, finance and many other areas. Given a dataset consisting of $\{\bX_i, Y_i \}_{i=1}^n$ where $Y_i\in \mR$ is the response and $\bX_i\in \mR^p$ is the covariate for the $i$th observation, the main interest is often to conduct a regression analysis to relate $Y$ to $\bX$,  the simplest model for which takes the linear form.

  An important assumption in linear regression is usually that the observations are all generated  from the same model.   In many applications, however,    the data collected often  contain contaminated or noisy observations due to a plethora of reasons. Those observations exerting great influence on statistical analysis, thus named influential points, can seriously distort all aspects of data analysis such as alter the estimation of the regression  coefficient and sway the outcome of statistical inference \citep{Draper:Smith:2014}. Thus, when influential points are present, fitting the model based on a clean data assumption leads to at best a very crude approximation to the model and at worst a completely wrong solution.  For fixed dimensional models, we refer the reader to \cite{Cook:1977,Belsley:etal:1980,Chatterjee:Hadi:1986,Imon:2005, Zhu:etal:2007,Zhu:etal:2012,Nurunnabi:etal:2014}, among many others.
  For high-dimensional models, \cite{Zhao:2013} found that influential observations could negatively impact many methods recently developed for dealing with high-dimensionality, such as Lasso for variable selection (Tibshirani, 1996) and SIS for variable screening (Fan and Lv, 2008).

  As a result, influence diagnosis has been long recognized as a central problem and  routinely recommended in statistical analysis. An entire line of research has been devoted to devising robust methods that are less prone to influential observations; See, for example, an excellent book on robust regression by \cite{Huber:2011}  when $p$ is fixed. 
  \cite{Wang:etal:2007} and \cite{Fan:etal:2014}, among others, devised robust methods for variable selection when heavy tailed noises are present, but no attempt was made to 
   to quantify the influence of individual points, which can often be the main question of interest in practice. 
 For multivariate data containing only $\bX_i$'s, \cite{Aggarwal:Yu:2001} proposed to find outliers in a high-dimensional space via projection, while \cite{Ro:etal:2015}  used a robust covariance matrix estimator for defining distance for detecting outliers.   \cite{She:Owen:2011}  is among the first to study outlier detection in regression. Focusing on the mean shift model for $p<n$ problems,  they did not show any theoretical guarantee for outlier dection.  It is also found that empirically She and Owen's method is outperformed by the approach proposed in this paper (Section 4).

 When $p$ is fixed,  there are many measures proposed for quantifying the influence of each observation, noticeably,
Cook's distance \citep{Cook:1977}, Studentized residuals \citep{Velleman:Welsch:1981:studentized_residual},  DFFITS \citep{Welsch:Kuh:1977:MIT,Belsley:etal:1980}, and  Welsch's distance \citep{Welsch:1982}. These measures have now been implemented in most statistical software such as R and SAS. %
Since these measures are all based on the ordinary least squares (OLS) estimation, they are not applicable to high-dimensional data.  On the other hand, despite its obvious importance, the problem of influence diagnosis in a high-dimensional setting has received little attention.  This is mainly due to the difficulty in establishing a coherent theoretical framework, even in a fixed-dimension setting, and lack of easily implementable procedures.
 \cite{Zhao:2013} appears to be the the first work on high-dimensional influence diagnosis. They proposed a  new high-dimensional influence measure named HIM based on marginal correlations and established its asymptotic properties. The asymptotic theory further permits the development of a multiple testing based  procedure for detecting influential points.

Similar to many fixed dimensional measures, HIM  is based on the idea of leave-one-out. That is, to quantify the influence of an observation, one compares a predefined measure evaluated on the whole dataset and the measure evaluated on a subset of the data leaving out the observation under investigation. Because of this, HIM is useful for detecting the presence of a single influential point.  In practice, however,  multiple influential observations are commonly encountered and it is not appropriate to apply a test for a single influential point sequentially in order to detect multiple ones.  On the other hand, detecting multiple influential observations  is  much more challenging, due to  the notorious ``masking" and ``swamping" effects \citep{Hadi:1993}. Specifically, masking  occurs when an influential point is not detected as influential, while swamping occurs when a non-influential point is classified as influential.  In the language of multiple testing, masking is the problem of getting false negatives and swamping is the problem of getting false positives. To handle  the masking and swamping effects in fixed dimensional models,
 many group deletion methods have been proposed  (Rousseeuw and Zomeren, 1990;    Hadi and Simonoff, 1993; Imon, 2005; Pan et al., 2000; Nurunnabi et al., 2014, Roberts et al., 2015). Dealing with these effects for high-dimensional data, however, is much  more challenging  and is currently an open problem.

The main aim of this paper is to propose a new procedure for detecting multiple influential points for high-dimensional data based on HIM.
 Via random group deletion, we propose a novel procedure named MIP, short for \underline{m}ultiple  \underline{i}nfluential \underline{p}oint detection for high-dimensional data.   Along the process, we propose two novel quantities named Max and Min statistics to assess the extremeness of each point when data are subsampled.
 Our theoretical studies show that these two statistics have complementary properties. The Min statistic is useful for overcoming the swamping effect but less effective for masked influential observations, while the Max statistic is well suited for  detecting   masked influential observations but is less effective in handling  the swamping effect.  Combining their advantages, we propose a  computationally  simple Min-Max  algorithm for obtaining   a clean subset of the data that contains no influential points {with high probability}. This clean set of data is then served as the benchmark for assessing the influence of other observations, which permits one to control the false discovery rate of influential points by using, for example, the Benjamini-Hochberg procedure  \citep{Benjamini:1995}.  Remarkably, the  theoretical properties of Max and Min statistics can be studied and are rigorously established in this paper.  We must point out that even for fixed-dimensional problems, there is a general lack of principled procedures for declaring significance of any defined influence measures. On the contrary, our proposed MIP procedure is the first theoretically justified method and for the more challenging high-dimensional setting.  

\begin{figure}[!bpt]
\centering
\caption{\label{Fig1} Influential point detection by using the Max (plot (a)) or Min (plot (b)) statistic.   In (a), identified influential points are colored in either red or blue, while in (b), identified influential points are colored in blue. MIP identifies the $7$ blue points as influential.}\vspace{-0.8cm}
\begin{subfigure}[b]{.45\textwidth}
%\centering
\includegraphics[width=\textwidth]{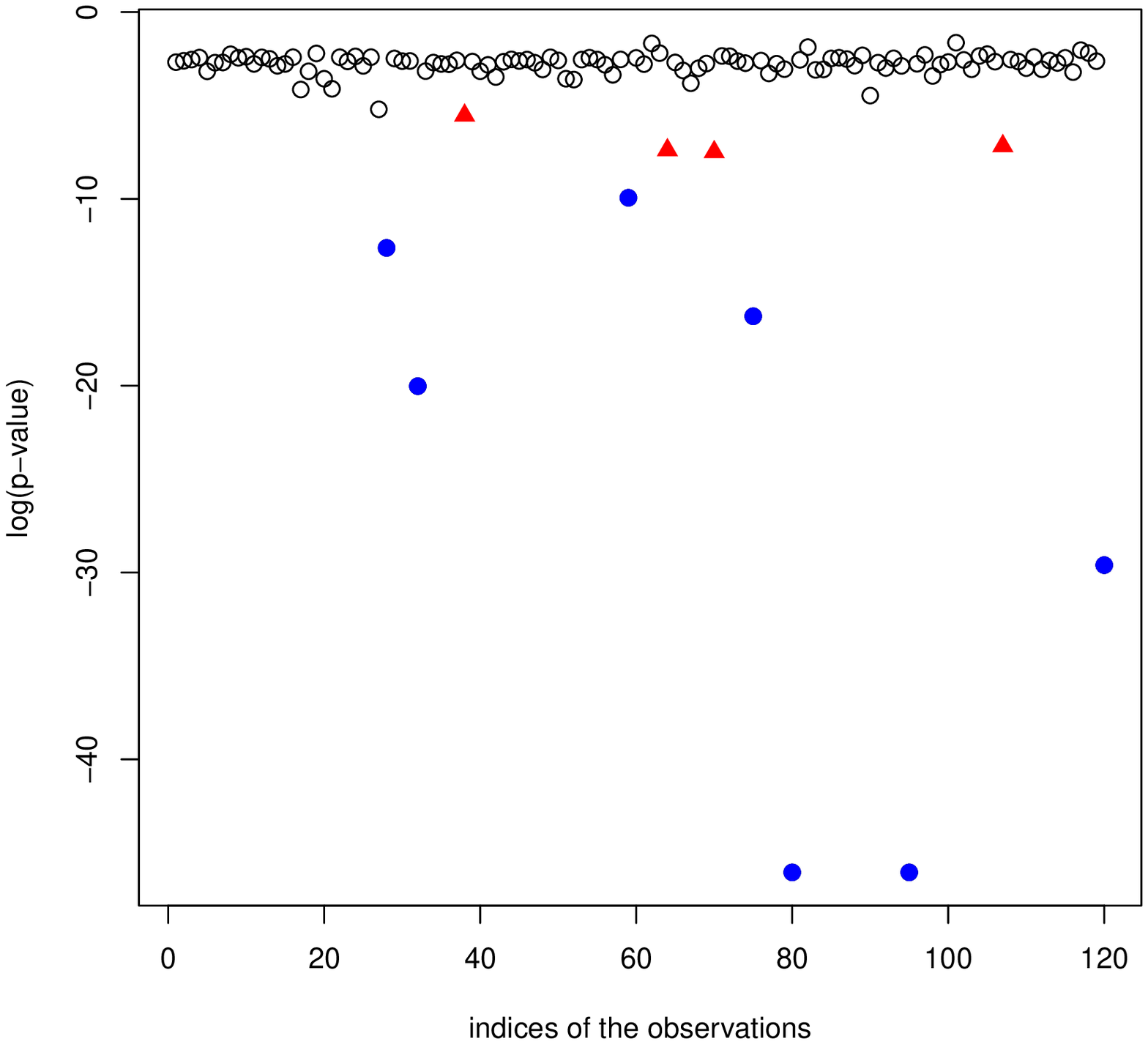}
\caption{log $p$-values by using the Max statistic. }
\end{subfigure}
\begin{subfigure}[b]{.45\textwidth}
%\centering
\includegraphics[width=\textwidth]{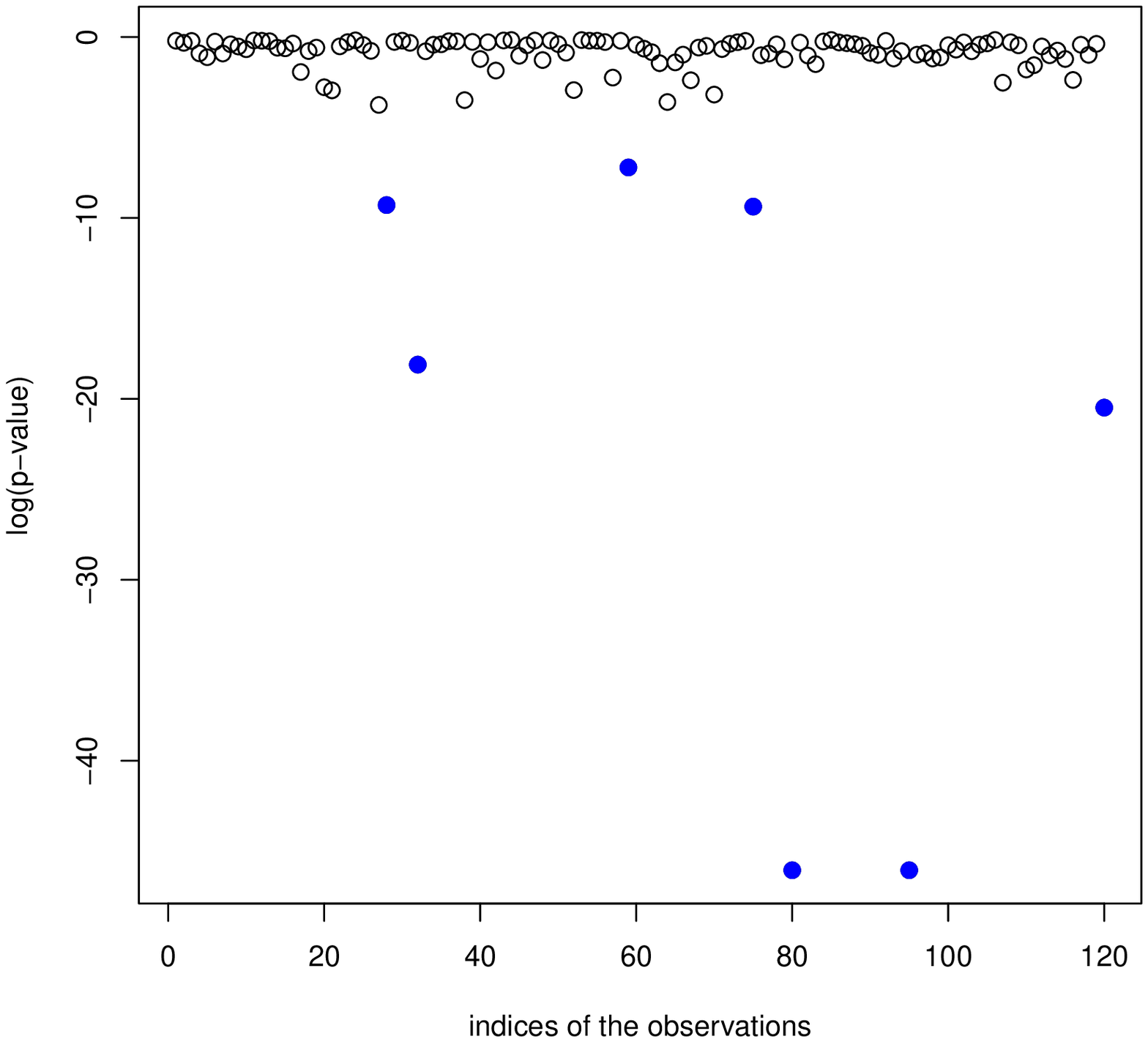}
\caption{log $p$-values by using the Min statistic. }
\end{subfigure}
\end{figure}

Before we proceed, we highlight the usefulness of the Max and Min statistics via an analysis of the microarry data in Section 4.3.  Figure \ref{Fig1} plots  the logarithms of the $p$-values associated with  the Max statistic in (a) and the Min statistic in (b) of the observations, respectively.
With a prespecified false discovery rate of $0.05$,  using the Min statistic, we identify a set of $7$ influential observations, represented as the blue points in plot (a) and (b). It is interesting that the MIP procedure combining the strengths of the two statistics identifies the same set of $7$ influential points. On the other hand, using the Max statistic, $4$ additional observations, represented as red  triangles in plot (a), are declared influential. These findings are consistent with our theory that  the Max statistic tends to identify  more influential observations, making it  more suitable for overcoming the masking effect, but  may suffer from the swamping effect.  On the other hand,  the fact that the Min statistic gives the same set of influential points as MIP in plot (b)   implies that there may not exist  any {masking} effect in this data.   Further analysis in Section 4.3 shows that the reduced data, obtained by removing the influential observations identified by MIP,  results in  a sparser model with a better fit, when Lasso is applied for model fitting.

The main flow of this paper is organized as follows.
In Section 2, we review the high-dimensional influence measure in \cite{Zhao:2013}. 
 In Section 3, based on the idea of random group deletion or leave-many-out,  we propose Max and Min statistics  for assessing extremeness and  establish their theoretical properties. The Max and Min statistics for a given point are the maximum and the minimum quantity, respectively, of the influence measures defined over randomly subsampled data. We show in Theorem \ref{Th2}  that,  surprisingly, when there is no influential point, these two statistics both follow  a $\chi^2(1)$ distribution.  When there are influential points, Theorem \ref{Th3} and Theorem \ref{Th4} show that for a non-influential point, its Max and Min statistics still follow a $\chi^2(1)$ distribution. Furthermore with the presence of influential points, Theorem \ref{Th3}  and \ref{Th4}  demonstrate that, under suitable conditions, the Max and Min statistics can identify the influential points with large probability. We then argue that  these two statistics are complementary in detecting influential observations and the Min-Max algorithm can suitably combine  their strengths.  Simulation results and data analysis, showing the competitive performance of MIP in comparison to HIM and the method of \cite{She:Owen:2011}, are presented in Section 4. In Section 5, we provide further discussions. All the proofs are relegated to the Appendix. An R package implemeting MIP, freely available on \url{http://www.warwick.ac.uk/chenleileng/research/} now, will be deposited onto CRAN.

Here are the notations used throughout the paper.
For any set $A$,  we write $|A|$ as its cardinality.
Let $S_{\inf}$ and $S^c_{\inf}$ be the set of the influential and non-influential  observations, respectively.
Denote by $\|v\|$   the $l_2$ norm of a vector $v\in \mR^m$. For any matrix $A=(a_{ij})\in \mR^{m\times n}$,     $\|A\|$ denote its  spectral norm, respectively. Finally, let $\|A\|_{\max}=\max\limits_{i,j}|a_{ij}|$ and we use $C$ to denote a generic constant that may change depending on the context.

%=============

\section{HIM, Masking and Swamping}

 \subsection{Review of HIM}
We first review the high-dimensional influence measure (HIM) in \cite{Zhao:2013} when $\min\{p, n \}\rightarrow \infty$.   Assume that the non-influential observations are $i.i.d.$ from the following model
\beq
\label{eq:model}
Y_i= \bX_i^\top\mathbf{\beta}+\varepsilon_i,~~i=1, ..., n,
\eeq
where $Y_i\in\mR$ is the response variable, $\bX_i=(X_{i1},\cdots,X_{ip})^\top\in\mR^p$ is the associated $p$-dimensional predictor vector, $\mathbf{\beta} \in \mR^p$ is the coefficient vector,
and $\varepsilon_i\in\mR$ is a   normally distributed random noise with $\cov(\bX_i,\varepsilon_i)=0$.
Denote $\mu_{y}= E(Y_i)$,  $\sigma_y=(\mathrm{var}(Y_{i}))^{1/2}$ and  $\mu_x=(\mu_{x1},\cdots,\mu_{xp})^\top=E(\bX_i)$,  $\sigma_{xj}=(\mathrm{var}(X_{ij}))^{1/2}, 1\le j\le p$.

The idea of HIM is to define the influence of a point by measuring its contribution to the average marginal correlation between the response and the predictors.  Specifically, define the marginal  correlation between variable $j$ and the response as $\rho_j = \mathrm{corr}(X_{ij}, Y_i)$. Given the data,  we can obtain its sample estimate as
$\hat{\rho}_j = \{\sum_{i=1}^{n} (X_{ij} - \hat{\mu}_{xj})(Y_i - \hat{\mu}_y)\}/\{n \hat{\sigma}_{xj} \hat{\sigma}_y \}$,
for $j = 1, \ldots, p$,
where $\hat{\mu}_{xj}, \hat{\mu}_y, \hat{\sigma}_{xj}$ and $\hat{\sigma}_y$ are the sample estimates of $\mu_{xj}, \mu_y, \sigma_{xj}$ and $\sigma_y$, respectively.    The sample marginal  correlation  with the $k$th observation removed is similarly defined as $\hat{\rho}_j^{(k)}$ for $1\le k\le n$.
 HIM then measures the influence of the $k$th observation by comparing the sample correlations with and without this observation, defined formally as
 \[\mathbb{D}_k=p^{-1}\sum_{j=1}^p \left (\hat\rho_j-\hat\rho_{j}^{(k)}\right )^2,~  1\le k\le n.\]
Intuitively, the larger $\mathbb{D}_k$ is, the more influential the corresponding observation is.
 When there is no influential point and $\min\{n,p\}\rightarrow \infty$, under mild conditions, it is proved that
$n^2\mathbb{D}_k \rightarrow \chi^2(1),$
where $\chi^2(1)$ is the chi-square distribution with one degrees of freedom.
Based on this result, we can formulate the problem of influential point detection as a multiple hypothesis testing problem where one tests $n$ hypotheses, one for each observation stating that the observation under invstigation is non-influential. Subsequently,
the Benjamini-Hochberg  procedure  \citep{Benjamini:1995} for multiple testing can be used to control the false discovery rate.

We now discuss why marginal correlation is attractive for defining influence.
Cook's distance and other classical influence measures rely on OLS which is infeasible in a high-dimensional setting whenever $p>n$.  Constrained versions of OLS such as Lasso might seem useful, but their properties are extremely difficult to establish if the leave-one-out scheme is to be employed for studying influence. Even with additional assumptions such as sparsity on $\beta$, it is unlikely that the difference between the estimates with all the data and all the data but one can be rigorously established.
On the other hand, an immediate advantage of using marginal correlation is that, as an ubiquitous quantity in statistics, it is well defined and more importantly tractable under this setting \citep{Zhao:2013}. Because of this, marginal correlation has also been used previously for other tasks such as variable screening \citep{Fan:2008}.

Next we discuss what we mean by influence by investigating what  points can be flagged up by HIM. First of all, if the covariance matrix of the covariates is diagonal, the marginal correlation vector $\rho=(\rho_1, \cdots, \rho_p)^\top$ is equivalent to $\beta$ at the population level. Thus in this case $\mathbb{D}_k$ can be loosely seen as a variant of the Cook's distance. In \cite{Zhao:2013}, HIM is further shown to be able to detect unusual points due to outlyingness in the response variable. More interestingly,  outlyingness in the covariates and points distorting the regression coefficient can also be detected by HIM, as we explain now.

Consider a simple mixture model in which $(\bX, Y)$ comes either from  $Y=\bX^\top \beta+\epsilon$ (Model 1) with probability $1-\theta$ or  $Y_{\inf}=\bX_{\inf}^\top\beta_{\inf}+\epsilon_{\inf}$ (Model 2) with probability $\theta$, where $\theta\in[0,1/2)$ is presumably small. With this setup, apparently, the aim of influence identification is to detect the points in Model 2.
For simplicity, assume that $\bX,\bX_{\inf}, \epsilon,$ and  $\epsilon_{\inf}$ all have mean zero.
Define
\[\rho_\theta:=E(\bX Y)=(1-\theta)E(\bX Y)+\theta E(\bX_{\inf}Y_{\inf})= (1-\theta) \cov(\bX) \beta+\theta \cov(\bX_{\inf})\beta_{\inf},\]
which is a function of $\theta$ whenever $\cov(\bX) \beta\not= \cov(\bX_{\inf})\beta_{\inf}$. By deleting one observation from the data as in HIM or multiple observations as in the MIP method, the empirical estimate $\hat\rho_\theta$ of $\rho_\theta$ changes as $\theta$ changes. This change can be fully exploited to identify influential points.
More specifically, when $\cov(\bX)=\cov(\bX_{\inf})=\bSigma$ but $\beta\ne \beta_{\inf}$, we have $\rho_\theta=\bSigma \beta_\theta$ where $\beta_\theta=(1-\theta)\beta+\theta\beta_{\inf}$.  There is a one-to-one mapping between $\rho_\theta$ and $\beta_\theta$.
      The change in marginal correlation $\rho_\theta$ indicates a change in   $\beta_\theta$ after  re-scaled by $\bSigma$. Finding observations that influence marginal correlation is, in some sense, equivalent to  identifying  those that influence the  regression coefficient. Furthermore, when there are abnormal points from covariates in that
   $\cov(\bX)\ne \cov(\bX_{\inf})$ but $\beta=\beta_{\inf}$, we can write $\rho_\theta=\bSigma_\theta\beta$ where $\bSigma_\theta=(1-\theta)\cov(\bX)+\theta\cov(\bX_{\inf})$. Again, there is  a one-to-one correspondence between $\rho_\theta$ and $\bSigma_\theta$. Identifying points that are abnormal in $\rho_\theta$ is equivalent to  finding points abnormal in the covariates.   In summary, the marginal correlation based measures can find influential points in the response, in the covariates, and in the coefficient, { and  HIM can be viewed as a screening method in this sense.}

\subsection{The effect of masking and swamping}
\begin{figure}[!hbpt]
\caption{\label{Fig2} Performance comparison between HIM and MIP. TPR: True positive rate; FPR: False positive rate. The nominal FPR is set at $\alpha=0.05$, {corresponding to the   horizontal dotted grey line.}}\vspace{-0.8cm}
\begin{subfigure}[b]{.5\textwidth}
\centering
\includegraphics[width=\textwidth]{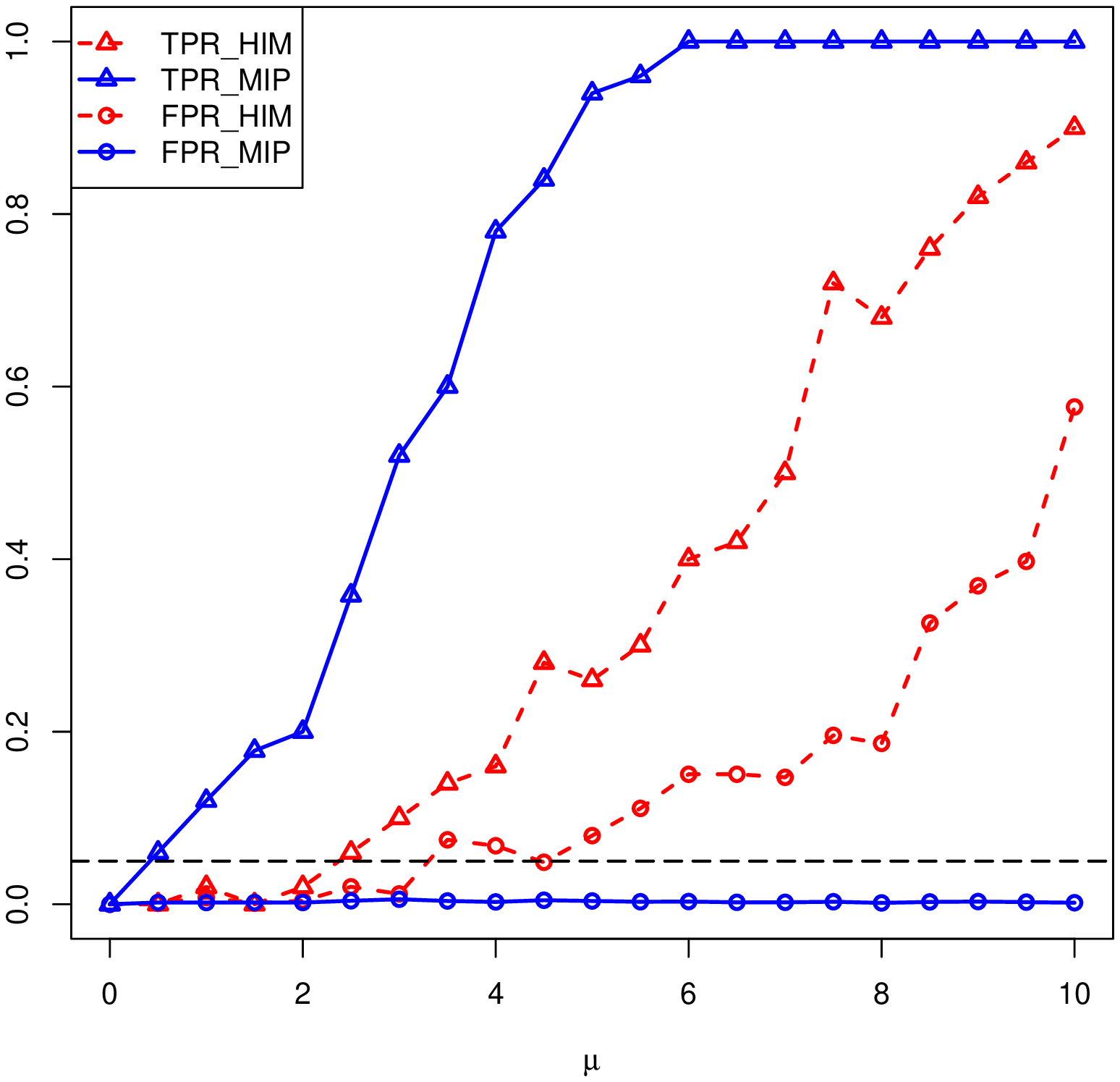}
\caption{Masking effect example (Example 1)}
\end{subfigure}
\begin{subfigure}[b]{.5\textwidth}
\centering
\includegraphics[width=\textwidth]{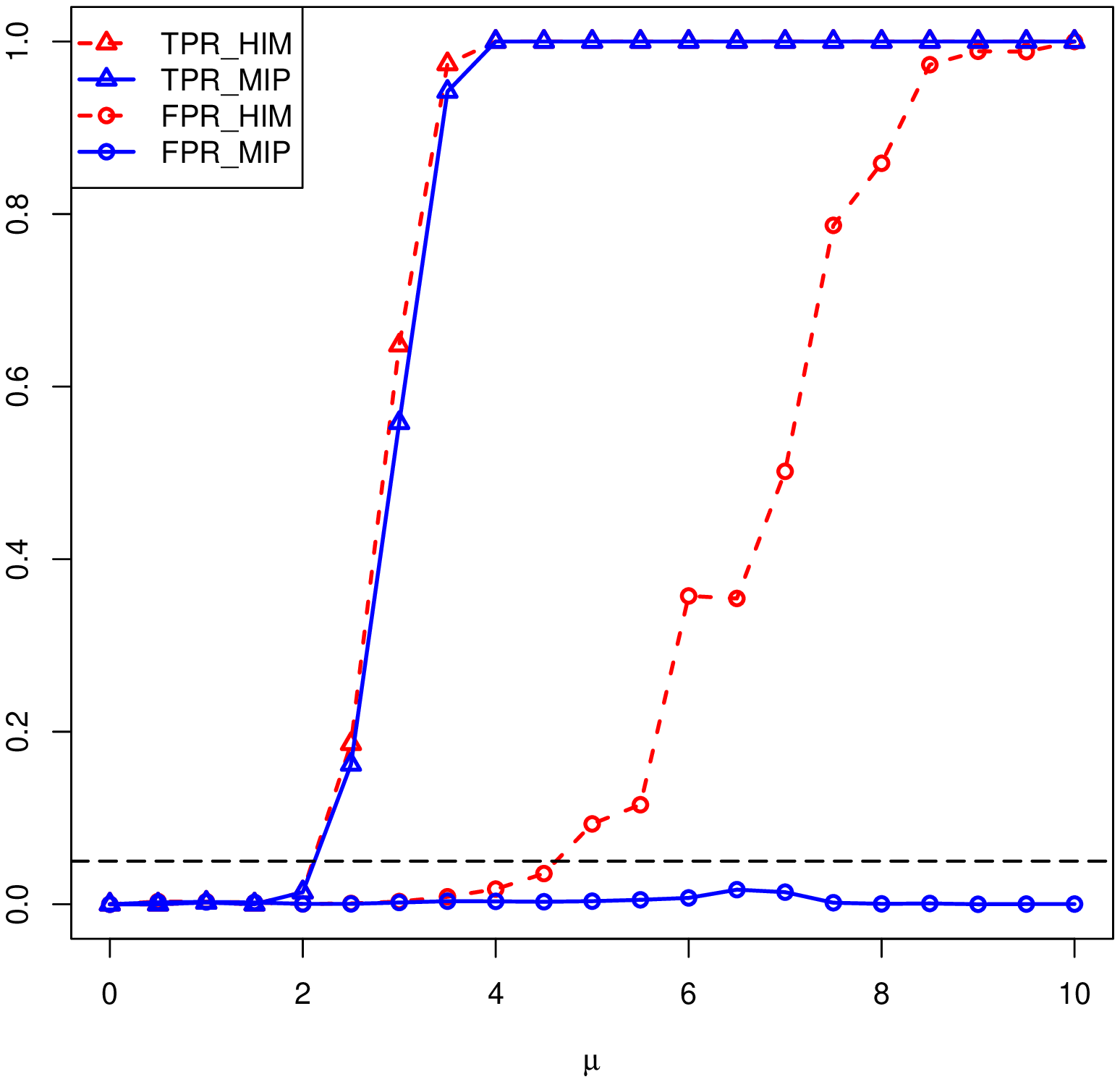}
\caption{Swamping  effect  example (Example 2)}
\end{subfigure}
\end{figure}

Since HIM is based on the leave-one-out idea, the derived $\chi^2(1)$ distribution is invalid whenever there are one or more influential points. That is, for a non-influential point , the presence of even one single influential point can distort the null distribution of its HIM value according to the definition above. Similarly, the presence of more than one influential point can distort  the HIM value of an influential point as well.
This is the manifestation of a more general difficulty of multiple influential point detection where the masking and swamping effects greatly hinder the usefulness of any leave-one-out procedures.  
To appreciate how masking and swamping effects negatively impact the performance of HIM,   we quickly look at Example 1 and 2 in Section 4.  The data are generated such that there exists a strong masking effect in Example 1  and  a strong swamping effect in Example 2. The magnitude of these effects depends on a parameter denoted as $\mu$ .    Figure \ref{Fig2} presents a comparison of  HIM in \cite{Zhao:2013} and the proposed MIP method  proposed in  this paper for detecting influence, when the nominal level used for declaring influential in  the Benjamini-Hochberg  procedure  is set at $\alpha=0.05$.

From plot (a) of  Figure \ref{Fig2}, we see that the  true positive rates (TPRs) of HIM are much lower than those of MIP; that is, HIM identifies much fewer influential points as influential and thus  suffers severely from  the masking effect. Meanwhile, the false positive rates (FPRs) of HIM are also much larger than the nominal level $\alpha=0.05$ especially when $\mu$ becomes large; that is, HIM identifies much more non-influential points as influential, meaning that HIM also suffers from the swamping effect. From plot (b), we see that HIM suffers from the swamping effect greatly, as the FPRs can be very close to 1 for large $\mu$. On the other hand, for both examples, the FPRs of the MIP procedure are controlled well below the nominal level while its TPRs are monotone functions of $\mu$ and eventually become one for large $\mu$.

%========
\section{A Random Group Deletion Procedure}

As discussed before, any measure based on the leave-one-out approach may be ineffective when there are multiple influential observations due to the  masking and swamping effects. Since  the number of influential observations is generally unknown in practice, it is natural to employ  a notion of leave-many-out or group deletion. Group deletion has also been used  for fixed dimensional problems in identifying multiple influential points \citep{Lawrence:1995,Imon:2005,Nurunnabi(2011),Nurunnabi:etal:2014, Roberts:etal:2015}, {where deletion is often  made according to  the magnitude of (studentized) residuals  or similar criteria and  a good estimate of $\beta$ is necessary.
However,  in the high dimensional setting considered in this paper, extending these methods   is challenging.

 For our random group deletion procedure,  the subsets are chosen with replacement uniformly at random. Thus, the marginal correlations based on these subsets can be seen as some kind of perturbations to the marginal correlations based on the whole sample. Their extremeness is summarized by two extremal statistics whose theoretically properties can be studied. Existing  group deletion procedures   are  not   employed in a way similar to how we define our statistics which are theoretically tractable.
 
}

Recall that    $S_{\inf}$ and $S_{\inf}^c$ denote the indices of influential and non-influential observations such that $S_{\inf}\cup S_{\inf}^c=\{1,\cdots, n\}$. Let  $|S_{\inf}|=n_{\inf}$ be the size of influential point set and $|S_{\inf}^c|=n-n_{\inf}$ be the number of non-influential points.
Write $Z_k=(\bX_k,Y_k), 1\le k\le n$ as the $k$th data point.
 For any fixed $k$, to check  whether $Z_k$ is influential or not, we draw uniformly at random with replacement some subsets   $A_1, \cdots, A_m\subset\{1, \cdots, n\}/ \{k \}$; that is, these subsets do not include $Z_k$. {The choice of $m$ will be discussed  in Section 3.3 and Section 4.}  Write   $|A_r|=n_{\mathrm{sub}}-1$ where $n_{\mathrm{sub}}=k_{\mathrm{sub}} n+1$ for some $k_{\mathrm{sub}}\in (0,1) $.  These subsets are repeatedly drawn in the hope that there exists some subset that contains no influential observations. If such a clean set can be found, then  the statistic associated with any non-influential point  has the $\chi^2(1)$ distribution as HIM.
 A conservative choice for  $k_{\mathrm{sub}}$ is $1/2$, because the number of non-influential points is usually larger than that of the influential points. Formally, we make the following assumption on $n_{\inf}$ and  $k_{\mathrm{sub}}$.
\begin{itemize}
   \item[(C1)]  Denote $\delta_{\inf,n}=n_{\inf}/n$ which is allowed to vary  with $n$.     Assume    $0\le \delta_{\inf,n}<1/2-\delta_1$  for some  $\delta_1>0$ independent of $n$.  We take      $k_{\mathrm{sub}}>\lim\sup\limits_n\delta_{\inf,n}+\delta_1$.
\end{itemize}
Assumption (C1) allows $\min\limits_n \delta_{\inf,n}\rightarrow0$.  For $1\le r\le m$, let  $B_r$ be  the subset of non-influential observations in $A_r$ and denote its size as $N_{B_r}=|B_r|$.      Under (C1), we have $\min\limits_{1\le r\le m}N_{B_r}>\delta_1n$, that is,  for any  subset $A_r$,  the number of non-influential observations does not vanish.

  For $1\le r\le m$, let  $A_r^{(+k)}=A_r\cup \{k\}$ which is of size $n_{\mathrm{sub}}$. For $Z_{k}$, we compute its influence measure  with respect to the $r$th random subset $A_r$ as
$$
\mD_{r,k}=p^{-1}\|\hat\brho_{A_r^{(+k)}}-\hat\brho_{A_r}\|^2, \quad  1\le r\le m,
$$
where $\hat\brho_{A_r}$ and $\hat\brho_{A_r^{(+k)}}$ denote   the estimate of $\brho$  based on observations in  $A_r$ and $A_r^{(+k)}$, respectively.  We are now ready to define the following two extreme statistics,
$$T_{\min,k}=\min\limits_{1\le r\le m}n_{\mathrm{sub}}^2\mD_{r,k}, \ \ \ \ \  \   T_{\max,k}=\max\limits_{1\le r\le m}n_{\mathrm{sub}}^2\mD_{r,k}.$$
We name them  the Min and Max statistic respectively as they measure the extremeness of the influence measures based on randomly sample data.  {Note that the statistics  defined here, using    Euclidean norm,  are   invariant to the  rotation of the covariates and to the scale translation  of the response.}

To establish  the asymptotic behaviours of $T_{\min,k}$ and $T_{\max,k}$, we first study the behaviour of a key quantity  $J_{\max,n}=\max\limits_{1\le r\le m}J_r$ in which $J_r$ is defined as
$$J_r= p^{-1}\sum\limits_{j=1}^p\|\frac{1}{N_{B_r}}\sum\limits_{t\in B_r }  \hat Y_t\hat X_{tj}\|^2=p^{-1}\|\frac{1}{N_{B_r}}\sum\limits_{t\in B_r }  \hat Y_t\hat \bX_{t}^\top\|^2,$$
where  $\hat Y_t=\hat\sigma_y^{-1}(Y_t-\hat\mu_y)$, $\hat\bX_t=\hat D_x^{-1}(\bX_t-\hat\mu_x), 1\le t\le n$, and $\hat D_x$ is  the estimate of  $D_x=\mathrm{diag}(\sigma_{x1},\cdots, \sigma_{xp})$, a diagonal matrix in $ \mR^{p\times p}$.
By definition, $J_r$ is the square of $\ell_2$ norm associated with the non-influential observations in $A_r$ only {and is therefore unknown}.
  Denote $\dot \bX_t=D_x^{-1}(\bX-\mu_x)$ as the population version of $\hat\bX_t$ and note that $\dot Y_t$ is  the population version of  $\hat Y_t$.  {Without loss of generality, we   assume    in  model (\ref{eq:model})    that   $\mu_y=\mu_x=0$ and $\sigma_y=\sigma_{xj}=1, 1\le j\le p$, respectively. Moreover,  we make the following assumptions. }

\begin{itemize}
  \item[{(C2)}] For $1\le j\le p, 1\le s\le q$, $\rho_{js}$ is constant and does not change as $p$
increases.
  \item[{(C3)}] For the covariance matrix of the covariates  $\Sigma= \cov(\bX_i)$ with  eigen-decomposition
$\Sigma =\sum\limits_{j=1}^p \lambda_ju_ju^\top_j$, we assume $l_p =\sum\limits_{j=1}^p \lambda_j^2 =O(p^r)$ for some $0 \le  r < 2$.
  \item[{(C4)}]The predictor $\bX_i$ follows a multivariate normal distribution and the
random noise $\varepsilon_i$ follows a multivariate  normal distribution with mean zero and an unknown variance.
\end{itemize}

{
\begin{itemize}
\item[(C5)]  Let  $(Q_{y},R_y)=((\hat\mu_y-\mu_y)/\sigma_y, \sigma_y/\hat\sigma_y-1)$, $S_{Qy}=\limsup\limits_{n\rightarrow\infty}E(n^{1/2} Q_y)^8$ and  $S_{Ry}=\limsup\limits_{n\rightarrow\infty}E(n^{1/2}R_y)^8$.
Assume that     $S_{Q_y}$ and $S_{R_y}$ are finite.
Furthermore, there exist  constants $0<K, C<\infty$, independent of $n$ and  $p$, such that for any $t>0$,
$$\max\limits_{1\le j\le p}P(|\hat\mu_{xj}-\mu_{xj}|>t/\sqrt{n})\le C\exp(-t^2/K),$$
$$\max\limits_{1\le j\le p}P(|\hat\sigma_{xj}/\sigma_{xj}-1|>t/\sqrt{n})\le C\exp(-\min(t/K,t^2/K^2)). $$
\end{itemize}
Assumptions (C2)--(C4) are also made in \cite{Zhao:2013}. Since it is  assumed that $\sigma_{xj}=1, 1\le j\le p$, we have $tr(\bSigma)=p$ and consequently  it holds that  $l_p\le p^2$ by Cauchy-Schwarz inequality. When $l_p=p^2$, $\bSigma$ is a degenerate  matrix with rank one and (C3) rules out this   case. On the other hand, (C3) applies when the largest eigenvalue of $ \bSigma $ is bounded.  Assumption   (C5) is similar to but stronger than  (C.4) of \cite{Zhao:2013}, where only eighth moments of $n^{1/2}(\hat\mu_{xj}-\mu_{xj})$ and $n^{1/2}(\hat\sigma_x/\sigma_x-1)$ are required.  In Assumption   (C5),  $n^{1/2}(\hat\mu_{xj}-\mu_{xj})$  is assumed to have sub-Gauassian tails and  $n^{1/2}(\hat\sigma_{xj}/\sigma_{xj}-1)$'s have  sub-exponential tails.   This assumption is satisfied for the sample mean and the sample variance under the normality of $(\bX_i,Y_i)$'s.
As alternatives to the sample estimates, robust estimates of $\mu_x,\mu_y$, $\sigma_{xj}$, and $\sigma_y$ can also be used in practice. For example, we can  estimate $\mu_{xj}$ and $\mu_{y}$  by the sample median and  $\sigma_{xj}$ by the median absolute deviation (MAD) estimator, respectively. These estimates  satisfy Assumption (C5) by noting  the normality   of $(\bX_i,Y_i)$'s.  These robust estimates are the quantities used in our numerical examples.

} We now quantify the magnitude of $J_{\max,n}$, the maximum effect of the non-influential points, which is a key quantity for establishing the asymptotic properties of the Min and Max statistics.

\bel\label{Lemma1}
Assume that the non-influential observations satisfy  (C2)-(C4) and that (C1) and (C5) hold.  Assume further $\xi_{n,p}=n^{-1/2}(\log p)(\log n)\log(np)\rightarrow 0$.
 Then for any $1\le m\le \infty$,
  $$J_{\max,n}=O_p(\xi_{n,p}+p^{-1}l_p^{1/2}).$$
 \eel
Obviously, $\xi_{n,p}\rightarrow0$ if $n^{-1/4+\epsilon_0}\log p\rightarrow0$ for some sufficiently small $\epsilon_0>0$.  Here the number of the subsamples $m$ is allowed to grow to $\infty$ to help us understand  the   approach as explained in the next section, although in practice we only need $m$ to be large. Based on  Lemma \ref{Lemma1}, we have the following conclusion when there is no influential observation.

\bet\label{Th2} Suppose that all observations are non-influential.  Under the assumptions of Lemma \ref{Lemma1}, it holds that, for any $1\le k\le n$,
$T_{\min,k}\rightarrow_d \chi^2(1)$ and $T_{\max,k}\rightarrow_d \chi^2(1).$
\eet

Theorem \ref{Th2} seems surprising at first glance, since we always have $T_{\min,k}\le T_{\max,k}$. An explanation is in place. It will be shown that $\mD_{r,k}$ can be decomposed into two parts. {The first part,  depending on  the quantity $E_k$ defined in the next paragraph, represents the effect of   the observation $Z_k$, and the second part  is  controlled by $J_{\max,n}$. Since $J_{\max,n}={o_p}(1)$ by Lemma \ref{Lemma1}, the asymptotic distributions of $T_{\min,k}$ and  $T_{\max,k}$ are  mainly determined by $E_k$. Thanks to  the blessing of dimensionality, we can show that $E_k$ asymptotically  has  a $\chi^2(1)$ distribution.}  From Theorem \ref{Th2}, when $T_{\max,k}$ or $T_{\min,k}$ is larger than $\chi^2_{1-\alpha}(1)$, the $(1-\alpha)100\%$ quantile of the $\chi^2(1)$ distribution, for some prespecified $\alpha$ such as $0.05$, {we declare that there exist outliers.}

Recall that $B_r$ is the set consisting of the indices of the non-influential observations in $A_r$.  Let $O_r=A_r\setminus B_r$ be its complment in $A_r$. For each $1\le r\le m$,    it is obvious that $O_r\subseteq S_{\inf}\setminus\{k\}$, the latter equal to  $S_{\inf}$ if $k\in S_{\inf}^c$.  Since $|A_r|=n_{\mathrm{sub}}-1=k_{\mathrm{sub}}n$,   similar  to the proof of Theorem \ref{Th2}, we have
\beqr\label{dec_D_rk}
n_{\mathrm{sub}}^2\mD_{r,k}&=&p^{-1}\|\hat\brho-\hat\brho^{(k)}\|^2
 =p^{-1}\|\frac{1}{n_{\mathrm{sub}}-1}\sum\limits_{t\ne k,t\in A_r }  \hat Y_t\hat \bX_{t}^\top-\hat Y_k\hat \bX_{k}^\top \|^2\nonumber\\
&=&p^{-1}\|\frac{1}{nk_{\mathrm{sub}}}\sum\limits_{t\in B_r }  \hat Y_t\hat \bX_{t}^\top+\frac{1}{nk_{\mathrm{sub}}}\sum\limits_{t\in O_r }\hat Y_t\hat \bX_{t}^\top-\hat Y_k\hat \bX_{k}^\top \|^2\nonumber\\
&:=&p^{-1}\|W_{non,k,r}+W_{\inf,k,r}-\hat Y_k\hat \bX_{k}^\top  \|^2,
\eeqr
where  $W_{\inf,k,r}=\sum\limits_{t\in O_r }\hat Y_t\hat \bX_{t}^\top/nk_{\mathrm{sub}}$ and $W_{non,k,r}=\sum\limits_{t\in B_r }  \hat Y_t\hat \bX_{t}^\top/nk_{\mathrm{sub}}$ are associated with influential  and non-influential observations, respectively.
  Define
  $$E_{k}=p^{-1}\|\hat Y_k\hat \bX_{k}^\top\|^2,$$
which   represents the effect of the $k$-th observation $Z_k$. Let %
$$F_{\min,k}=\min\limits_{1\le r\le m}p^{-1}\|W_{\inf,k,r}\|^2 ~~\ \mbox{and} \ \   F_{\max,k}=\max\limits_{1\le r\le m}p^{-1}\|W_{\inf,k,r}\|^2$$
 quantify the maximum and minimum  joint effect of the influential observations, respectively.
The asymptotic behavior of $T_{\max,k}$ and $ T_{\min,k}$  depends  on  the magnitude  of $E_k$, $F_{\min,k}$ and $F_{\max,k}$ when     multiple influential observations are present. See Theorem \ref{Th3} in Section 3.1 and Theorem \ref{Th4} in Section 3.2. We state the properties of $T_{\max,k}$ and $ T_{\min,k}$ separately.

%===========
\subsection{Max statistic $T_{\max,k}$ for the $k$th point}
In Theorem \ref{Th2}, we derive the null distribution of $T_{\max,k}$ and $T_{\min,k}$ when there is no influential point.
We now study $T_{\max,k}$ when there are influential observations and develop the corresponding  detection procedure.  %
  Recall  $n_{\inf}=n\delta_{\inf,n}$ and  $k_{\mathrm{sub}}>0$ in (C1). Denote   $\delta_{\inf,n}/k_{\mathrm{sub}}=R_{\inf}$, the ratio of $|S_{\inf}|$ over $|A_r|$,   and let  $d_{S}=\max\limits_{t\in S} E_t$, for any $S\subseteq S_{\inf}$. Simple calculation in the proof of Theorem \ref{Th3} shows   $F_{\max,k}\le \delta_{\inf,n}^2d_{S_{\inf}}$. We have the following results for $T_{\max,k}$.
\bet\label{Th3}
Under  the assumptions of Lemma \ref{Lemma1}, when there are influential observations, the following two conclusions hold.
\begin{itemize}
  \item[(i)]  Suppose further $F_{\max,k}\rightarrow 0$.  If observation $k$ is non-influential, that is, $k\in S_{\inf}^c$, then both $T_{\min,k}$ and $T_{\max,k}$ converge to  $\chi^2(1)$ in distribution.
  \item[(ii)]  For an influential point $k\in S_{\inf}$,  if
  \[ \mbox{Max-Unmask Condition}: E_{k}^{1/2}>\left(\chi^2_{1-\alpha}(1)\right)^{1/2}+F_{\min,k}^{1/2}\]
   holds for some small prespecified $\alpha>0$ where $\chi^2_{1-\alpha}(1)$ is the $100(1-\alpha)\%$ quantile of a $\chi^2(1)
$ distribution,        then $P(T_{\max,k}>\chi^2_{1-\alpha}(1))\rightarrow 1$.  In addition, it holds that  $F_{\min,k}<a_0^2<\infty$ for some $a_0>0$.
\end{itemize}
\eet
Under  the condition in $(i)$, for any non-influential observation $Z_k$, the asymptotic distributions  of  $T_{\min,k}$ and $T_{\max,k}$ are   the same as those in  Theorem \ref{Th2}. That is, the distributions of the Min and Max statistics of a non-influential observation are   not affected by the presence of influential observations. As such, a non-influential point can be identified as non-influential with high probability. That is, the swamping effect can be overcome under the condition in $(i)$.   Since  $F_{\max,k}\le \delta_{\inf,n}^2d_{S_{\inf}}$,  a sufficient condition for $F_{\max,k}\rightarrow0$ is that   $\delta_{\inf,n}^2d_{S_{\inf}}\rightarrow 0$, which holds if   $d_{S_{\inf}}<C<\infty$ and $\delta_{\inf,n}\rightarrow 0$.
 This condition  might be violated, however, if  $\delta_{\inf,n}$ does not vanish or   some influential observations have large values in terms of  $E_t$.  This condition implies that   deleting  points with large values in $E_t$ is helpful to  alleviate the swamping effect.

For an influential observation $Z_k$,  the Max-Unmask condition in $(ii)$ gives the requirement on its signal strength  for it  to be identified as influential.  As  $a_0$ decreases, the condition becomes weaker and easier to  be satisfied, and $Z_k$ is easier to be  detected.  This provides opportunity to identify  the  influential observations that are masked by others, as long as  we can make $a_0$ small enough.
In fact, as argued below, $a_0$ can be very small if $m$ is sufficiently large.

Now, we discuss the upper bound $a_0$ in $(ii)$ of Theorem  \ref{Th3}.
Recall that  $O_r$ denotes the indices of  the influential observations in $A_r$ and
 note $|O_r|\le n_{\inf}$. Then we have
  $$
  F_{\min,k}=p^{-1}\min\limits_{1\le r\le m}\|W_{\inf,k,r}\|^2\le  \min\limits_{1\le r\le m}\left[\left(\frac{|O_r|}{nk_{\mathrm{sub}}}\right)^2\max\limits_{t\in O_r} E_{t}\right] .$$
 Define $N_{O,m}=\min\limits_{1\le r\le m}|O_r|$. By allowing $m=\infty$, it is easy to see that   $N_{O,m}$  is a  decreasing function of $m$ with $\lim\limits_m N_{O,m}=0$, since there  are many subsets  $A_r$ that  contain no influential observations under assumption (C1), i.e. $|O_r|=0$. Therefore, $\lim\limits_m F_{\min,k}=0$. Of course, in practice $m=\infty$ is not achievable.
  Assume further  $d_{S_{\inf}}=\max\limits_{t\in S_{\inf}} E_t<C<\infty$. Then $F_{\min,k}\le C(N_{O,m}/(nk_{\mathrm{sub}}))^2$, which will be  small for large $m$ and $n$.
If $d_{S_{\inf}}$ is unbounded but $d_{S_{\inf}}/(nk_{\mathrm{sub}})^{2-\delta}<C<\infty$ for some $0<\delta<1$,   we  have $F_{\min,k} \le CN_{O,m}^2/(nk_{\mathrm{sub}})^\delta $, which converges to 0, as $m,n\rightarrow\infty$.  Generally, when $m$ and $n$ are large,   $a_0$  will be  small under some mild conditions.
 Therefore, $T_{\max,k}$ has advantages in overcoming  the masking effect if $m$ is large.

We formally formulate a multiple testing problem to test the influentialness of individual observations with $n$ null hypotheses $H_{0k}: Z_k$ is non-influential, $1\le k\le n$.
By $(ii)$ of Theorem \ref{Th3} and the above discussions, we can estimate the set of the    influential observations as
$$\hat S_{\max}=\{k: p_{\max,k}<q_k, 1\le k\le n\},$$
where $p_{\max,k}=P(\chi^2(1)>T_{\max,k})$ is the $p$-value under $H_{0k}$ and $q_k$'s  are determined by  the  specific  procedure used to control the error rate.  Here  $q_k$'s  can be independent of $k$, if we aim to control the familywise error rate by  the Bonferroni test. Alternatively, $q_k$'s can depend on $k$, if we want to control the false discovery rate (FDR) at level  $\alpha_0$. For example,  for the procedure in \cite{Benjamini:1995},  $q_k$ can be  taken as the largest $p_{\max,(k)}$ such that $p_{\max,(k)}\le k\alpha_0/n$, where $p_{\max,(1)}\le p_{\max,(2)}\le \cdots\le p_{\max,(n)}$ are the ordered $p_{\max,k}$'s. We now state the theory of using the Benjamini-Hochberg procedure and will use it later for numerical illustration, although other procedures developed for controlling FDR can also be used.
\bep\label{Proposition2}
Suppose that  the Benjamini-Hochberg procedure is used to control FDR at level $\alpha_0$.  If the Max-Unmasking  condition in  $(ii)$ of Theorem \ref{Th3}  holds with $\alpha<\delta_{\inf,n}\alpha_0$   but with $F_{\min,k}$ replaced by the constant  $a_0^2$ defined there, then under the conditions in Lemma \ref{Lemma1},  we have $P(\hat S_{\max}\supseteq S_{\inf})\rightarrow 1$.
\eep

Note that  $a_0$  discussed further after  Theorem \ref{Th3} is independent of $k$.
  Proposition \ref{Proposition2} shows that all the influential points will be identified as influential with high probability. That is,
  the true positive rate is well controlled.  In addition, if $\delta_{\inf,n}d_{S_{\inf}}\rightarrow0$, by   $(i)$ in Theorem \ref{Th3},  there will be no swamping effect and then the statistic $T_{\max,k}$  under  $H_{0k}$  follows $\chi^2(1)$ distribution.  Let $FPR(\hat S_{\max})=|\hat S_{\max}\cap S^c_{\inf}|/|S_{\inf}^c|$ be the estimated FPR.
 When  the Benjamini-Hochberg  procedure  is applied and there is no swamping effect,   $FPR(\hat S_{\max})$ will be  controlled.  However,   the condition $\delta_{\inf,n}d_{S_{\inf}}\rightarrow0$ is strong and it  may fail if $\delta_{\inf,n}$ does not converge to zero. In this case, FPR may be out of control.

To summarize, the  detection procedure based on the Max statistic   $T_{\max,k}$ is effective  in overcoming the masking effect, but  it  is  somewhat aggressive in that the FPR may not be  controlled well without strong conditions.
On the other hand, we point out that the procedure based on $T_{\max,k}$
is computationally efficient, compared with that based  on $T_{\min,k}$ below.

%========
\subsection{Min statistic $T_{min,k}$ for the $k$th point}
 We have argued that the statistic  $T_{\min,k}$ is  effective in alleviating the swamping effect.
 We formally state this in the following theorem.
\bet\label{Th4}
 Under the assumptions of Lemma \ref{Lemma1}, the following two conclusions hold.
\begin{itemize}
  \item [(i)] Assume $F_{\min,k}\rightarrow 0$. For any non-influential point $k\in S_{\inf}^c$, it holds that  $T_{\min,k}\rightarrow_d \chi^2(1)$.
  \item  [(ii)] For any influential $Z_k$, if
  \[ \mbox{Min-Unmask~Condition}:  E_k^{1/2}> F_{\max,k}^{1/2}+(\chi^2_{1-\alpha}(1))^{1/2}\] holds,  then
 $P(T_{\min,k}>\chi^2_{1-\alpha}(1))\rightarrow 1$,  where $\alpha>0$ is a small constant.
\end{itemize}
\eet

 Compared with $(i)$ of Theorem \ref{Th3} where $F_{\max,k}\rightarrow 0$ is required, the condition in  $(i)$ of  Theorem \ref{Th4} is much weaker. As discussed in Section 3.1,  $F_{\min,k}\rightarrow0$ when $\min\{m,n\}\rightarrow \infty$.   Therefore,  the statistic  $T_{\min,k}$ is less sensitive  to   the swamping effect.  On the other hand, $F_{\max,k}$ is involved in the Min-Unmask Condition in $(ii)$, which is much stronger than the Max-Unmask Condition in $(ii)$ of Theorem \ref{Th3}.   That is, an influential observation $Z_k$  will not be identified as influential unless its signal is very strong.  Thus, the Min statistic is efficient in preventing the swamping effect  but  may be conservative for identifying influential points.  Combining with the result in Section 3.1 that the Max statistic $T_{\max,k}$ is effective in overcoming the   masking effect but is aggressive,  we conclude that the  Max statistic $T_{\max,k}$ and the Min statistic $T_{\min,k}$ are complementary to each other.

If the Min-Unmask Condition holds for all $k\in S_{\inf}$ simultaneously, then  $Z_k$ with $k\in S_{\inf}$ will be detected  correctly,  when  certain error control procedure is used. For example,  similar to Proposition \ref{Proposition2}, with $\alpha=\delta_{\inf,n}\alpha_0$, one can show that the Benjamini-Hochberg procedure can correctly detect the influential observations.  However,  the Min-Unmask Condition  is very strong and may not be satisfied for  all $k\in S_{\inf}$ simultaneously.  We provide a sufficient condition for this condition to hold. Without loss of generality,  we assume $S_{\inf}=\{1,\cdots, n_{\inf}\}$ and  write $E_{(1)}\ge E_{(2)}\ge\cdots\ge E_{(n_{\inf})}$  ranking $E_{i}, 1\le i\le  n_{\inf},$ in a decreasing order.

 \bep\label{Prop3}
   If $E_{(n_{\inf})}^{1/2}>R_{\inf} E_{(1)}^{1/2}+(\chi^2_{1-\alpha}(1))^{1/2}$, then the Min-Unmask condition holds simultaneously for all the influential points $k\in S_{\inf}$.
  \eep

The condition in Proposition \ref{Prop3} is strong. When $\delta_{\inf,n}>0$ and $E_{(1)}$ is large, Proposition \ref{Prop3} needs $E_{(n_{\inf})}$  not to be too small but this condition may be violated easily.  A remedy is to sequentially remove the influential observations that have been detected so far and then apply  the detecting procedure recursively on the remaining data, as we explain below.

{\color{black}To simplify the description, we introduce some notations. For any subset $U\subseteq\{1,\cdots,n\}$ with cardinality $n_U=|U|$ and any observation  $Z_{k'}$ with $k'\in U$, we can draw at random with replacement subsets $A_{1,U},\cdots, A_{m,U}\subset U\setminus\{k'\}$, with the same cardinality $n_{\mathrm{sub},U}$, where $n_{\mathrm{sub},U}< n_U$.  Similar to  $T_{\min,k}$,  we define $T_{\min}(U,Z_{k'})=\min\limits_{1\le r\le m}n_{\mathrm{sub},U}^2D_{r,k',U}$, where   $D_{r,k',U}=p^{-1}\|\hat\brho_{A_{r,U}^{(+k')}}-\hat\brho_{A_{r,U}}\|^2$.   Denote by  $B_{r,U}$  the indices of non-influential observations in $A_{r,U}$ and let  $O_{r,U}=A_{r,U}\setminus B_{r,U}$, $1\le r\le m$. Let  $k_{\mathrm{sub},U}$ be  such that  $n_{\mathrm{sub},U}=n_Uk_{\mathrm{sub},U}+1$. Then  similar to $F_{\min,k}$, we  define $F_{\min}(U,Z_{k'})=\min\limits_{1\le r\le m} p^{-1}\|\sum\limits_{t\in O_{r,U}}\hat Y_t\hat\bX_t^T/(n_Uk_{\mathrm{sub},U})\|^2$, which denotes  the minimum of the joint effect of influential observations with indices  in $U$. And similar to $F_{\max,k}$, one can define $F_{\max}(U, Z_{k'})$.  Obviously, when $U=\{1,\cdots, n\}$, $T_{\min}(U,Z_{k'})$, $F_{\min}(U,Z_{k'})$ and  $F_{\max}(U,Z_{k'})$ are exactly the same as  $T_{\min,k'}$,   $F_{\min, k'}$ and $F_{\max, k'}$, respectively.
}

  Generally, suppose that  $E_{(i)}$'s can be separated into several groups in successive order, that is, $G_{j}=\{E_{(m_{j-1}+1)},\cdots, E_{(m_j)}\},j=1\cdots,\tau$, such that $0=m_0<m_1<\cdots<m_\tau=n_{\inf}$. Denote $I_j=\{(m_{j-1}+1),\cdots,(m_{j})\}, 1\le j\le \tau$.
   Let $M_0=S_{\inf}$, $M_j=M_{j-1}\setminus I_j$ and  $U_{j}=M_{j-1}\cup S_{\inf}^c$, $1\le j\le \tau $.
 For simplicity, we assume  that    $n_{\mathrm{sub},U_j}$'s are independent of $j$, denoted still as $n_{\mathrm{sub}}$,   and that the sufficient condition in {\color{black} Proposition  \ref{Prop3}}  holds for group $G_j$, that is,
 \beq\label{gMin-Unmaks}
 E_{(m_j)}^{1/2}>R_{\inf} E_{(m_{j-1}+1)}^{1/2}+(\chi^2_{1-\alpha}(1))^{1/2},\ \ 1\le j\le \tau,
 \eeq
  which is referred to as gMin-Unmask Condition  for simplicity.   Then, {\color{black} similarly to the argument of Proposition \ref{Prop3}, we see that  Min-Unmask Condition holds simultaneously  for any $Z_k, k\in I_j$  on the data set $\{Z_i, i\in U_{j}\}$, that is, $E_k^{1/2}> F_{\max}(U_j,Z_k)^{1/2}+(\chi^2_{1-\alpha}(1))^{1/2}$}. Consequently   $T_{\min}(U_j, Z_k)$ with  $Z_k\in I_j$ will be large than $\chi^2_{1-\alpha}(1)$ with high probability.  If influential observations in $I_{1},\cdots, I_{j-1}$ are detected correctly and removed sequentially, the influential observations in group $I_j$ can be detected successfully with high probability.   We remark that the  gUnmask-condition  is much weaker than the condition in Proposition \ref{Prop3}.

This motivates us to consider the following multi-round procedure.
 Define the set of  influential observations identified in  the $j$th round as
 $$\hat S_{\min}^j=\{k: P(\chi^2(1)>T_{\min}(\hat U_{j},Z_k))<q_k, Z_k\in \hat U_{j}\},$$
 where $q_k$ depends on the specific procedure used, similar to the discussion in Section 3.1, $\hat U_j=\hat U_{j-1}\setminus \hat S_{\min}^{j-1}$ with $\hat U_0=\{1,\cdots, n\}$, and $\hat S_{\min}^{0}=\emptyset$.   Finally, we can estimate $S_{\inf}$ by $\hat{\mathbb{S}}_{\tau'}=\cup_{j=1}^{\tau'} \hat S_{\min}^j$, where $\tau'$ is such that $\hat S_{\min}^{\tau'+1}=\emptyset$.
 Let $FPR(\hat{\mathbb{S}}_{\tau'})$ be the false positive rate associated with estimate $\hat{\mathbb{S}}_{\tau'}$.
\bep\label{Proposition3}
Suppose that  (C1) holds and that  FDR  is controlled at level $\alpha_0$ in each round. Then  ${E}(FPR(\hat{\mathbb{S}}_{\tau'}))\le \frac{\alpha_0}{1-\alpha_0}$.
\eep

Although the above iterative procedure can improve the performance of $T_{\min,k}$ to overcome the masking effect, requiring only  weaker gMin-Unmask Condition in (\ref{gMin-Unmaks}),   the computation of  this procedure will be more costly if the number of rounds $\tau'$ is large. On the other hand, the gMin-Unmask Condition will be easier to satisfy for larger $\tau'$.  Theoretically, $\tau'$ can be as large as $n_{\inf}$, where  gMin-Unmask Condition in (\ref{gMin-Unmaks}) becomes $F_{\min}=\min\limits_{t\in S_{\inf}} E_t>\chi^2_{1-\alpha}(1)/(1-R_{\inf})^2$  by noting that $E_{(m_i)}=E_{(m_{j-1}+1)}$, which is much weaker than the condition in Proposition \ref{Prop3}.   However, larger $\tau'$ demands more intensive  computing.
If an early stopping strategy is adopted, it may still suffer from  the masking effect.

As a quick summary, the test statistic $T_{\max,k}$ is more efficient  in dealing with the masking effect, because the strength of the influential observations required by  $T_{\max,k}$  in (ii) of Theorem \ref{Th3} is  much weaker than gMin-Unmask Condition (\ref{gMin-Unmaks}) required by $T_{\min,k}$, when $m$ is large.   Moreover, any procedure based on  $T_{\max,k}$ is computationally efficient,  identifying  the influential observations in just one round.
  However,  $T_{\max,k}$  may suffer from the swamping effect if the strong condition  $(i)$ of Theorem \ref{Th3} is violated. On the other hand, the estimate $\hat{\mathbb{S}}_{\tau'}$  based on the statistic $T_{\min,k}$ can maintain good FPR at the expense of more intensive computation.  Taking  advantages of both statistics, we propose the following   computationally efficient  Min-Max-Checking algorithm for identifying  {with high probability} a clean set that contains no influential points and can serve as the benchmark for assessing the influence of other points.

\subsection{Min-Max-Checking algorithm }
We propose the following algorithm to combine the strengths of the Max and Min statistics.

\begin{center}
{\bf Min-Max  algorithm  for estimating a clean set}
\begin{itemize}
 \item[]\quad {\bf Initialization}.  Let   $S_{total}=\{1,\cdots,n\}$ and  fix $c=1/2$. Repeat steps 1 and 2 until stop.
     \begin{itemize}
        \item[1.]{\bf Min-Step}.   For the data indices in $S_{total}$, compute $\hat M=\{k: P(\chi^2(1) >T_{\min,k})<\alpha_k, 1\le k\le n\}$.
            Alternatively we may simply take $\hat M$   as the set of indices with  the first $l_0$ smallest $p$-value for some small number $l_0$.
            Update $S_{total}\rightarrow S_{total}\setminus \hat M$.
        \item[2.]  {\bf Max-Step}.  Estimate $\hat S_{\max}$  as in Section 3.1 based on observations in $S_{total}$ and denote its complement $\hat S_{\max}^c$ as an estimate of the clean set. If $|\hat S_{\max}^c|\ge cn$, then stop; otherwise,  go to Min-Step.
     \end{itemize}
\end{itemize}
\end{center}
This algorithm identifies  {with high probability} a clean dataset containing no influential points with cardinality at least $n/2$ by successively removing potential influential points.
Here $\alpha_k$ is specified by  the procedure that controls the error rate, and can be   determined in the same way as $q_k$ in Section 3.1. The main rational of this algorithm is, as argued, that the Max statistic $T_{\max,k}$ is aggressive in declaring influential while Min statistic $T_{\min,k}$ is conservative.  We first run a Min-Step to eliminate those    influential observations with strong strength to alleviate the swamping effect.  Combined with the efficiency of $T_{\max,k}$ in overcoming   the masking effect,  it is highly possible to  obtain a  clean set with a large size in one iteration. If the clean set  is not large enough,  we run the Min-Step  again to remove further influential observations with strong strength.  In our numerical study, we find that this algorithm is computationally very efficient, usually stops in 1 or 2 rounds.

With some abuse of notations, write $\mathcal{S}_c$  as the final clean set  obtained by the Min-Max algorithm.
Then its supplement, written as $\mathcal{S}=\{1,\cdots,n\}\setminus \mathcal{S}_c$, is an estimate of the set which contains all potential influential observations. However, $\mathcal{S}$ may still contain non-influential observations  as the procedure for obtaining a clean set only aims to find a subset of the non-influential points.
A further step to check whether any point in $\mathcal{S}$ is truly influential if necessary. This step, however, is easy since we have now a clean dataset. We now outline the exact procedure.
For any $Z_i, i\in \mathcal{S}$, consider the data with indices in $\mathcal{S}_c$ and  $\mathcal{S}_c^{(i)}=\mathcal{S}_c\cup \{i\}$, respectively. We then compute  statistic $\mD_{i}$ as in Section 2 where $\hat\brho$ and $\hat\brho^{(i)}$ are computed on data set $\mathcal{S}_c$ and  $\mathcal{S}_c^{(i)}$, respectively.
 Since $\mathcal{S}_c$ is a good estimate of   the clean data containing no influential point,  this  leave-one-out approach will be effective for testing multiple null hypotheses in the form of $H_{0i}: Z_i\ \mbox{ is non-influential}, i\in  \mathcal{S}$.  If $\mathcal{S}_c$ is good, according to the results in HIM, $n_c^2\mD_i$ will follow $\chi^2(1)$ distribution under $H_{0i}$ by Theorem 1 of \cite{Zhao:2013}, where $n_c=|\mathcal{S}_c|+1$.
 The Benjamini-Hochberg procedure can then be applied to control FDR. Those whose corresponding hypotheses are rejected by the FDR procedure can be labeled as influential observations.  The algorithm for detecting multiple influential observations,  called {Min-Max-Checking} algorithm, is summarized as follows.
\begin{center}
{\bf Min-Max-Checking  algorithm }
\begin{itemize}
 \item[(1)] Estimate a clean subset $\mathcal{S}_c$ by the Min-Max algorithm;
\item[(2)] Check for each $k \in \mathcal{S}=\{1,\cdots,n\}\setminus \mathcal{S}_c$ whether the $k$th observation is influential.
\end{itemize}
\end{center}

\section{Simulation and Data Analysis}

 We evaluate the performance of MIP for detecting multiple influential points and compare it to HIM whenever possible.
  Throughout the simulation study, we set the sample size as $n=100$ and the number of predictors as $p=1000$.
 We generate $n$ observations from
 \beq\label{Model1}
 Y_i=\bX_i^\top\beta+\varepsilon_i, \quad 1\le i\le n,
 \eeq
 where   $\bX_i=(X_{i1},\cdots,X_{ip})^\top\in \mR^p$, $\beta=(\beta_1,\cdots, \beta_p)^\top\in \mR^{ p}$.  We then replace the first $n_{\inf}=10$ points in $\{(\bX_i, Y_i), i=1,\cdots, n\}$ by $\mZ^{\inf}=\{(\bX^{\inf}_i, Y^{\inf}_i), i=1,\cdots, n_{\inf}\}$ which are generated differently. The resulting dataset denoted as $\mZ_n$ thus may contain $10$ influential points. For \eqref{Model1},
 we set  $\varepsilon_i\sim N(0,1)$ and $\bX_i\sim N(0, \Sigma)$ where $(\Sigma)_{ij}=0.4^{|i-j|}$.
The coefficient  $\beta$ and how $\mZ^{\inf}$ is generated are specified below.

We evaluate performance by assessing the success in identifying influential and non-influential points, the accuracy in estimating $\beta$ in Model \eqref{Model1}, and the success in identifying the support of $\beta$. Let $S_{\inf}$ be the index set of the influential points and  $\hat S_{\inf}$ as  its estimate either by HIM or MIP.
We first compute $TPR_{\inf}$, the true positive rate for influential observation detection, and $FPR_{\inf}$,  the false positive rate  for detection. That is, $TPR_{\inf}=|\hat S_{\inf}\cap S_{\inf}|/n_{\inf}$ and  $FPR_{\inf}=|S_{\inf}^c\cap \hat S_{\inf}|/(n-n_{\inf})$. Denoting $FNR_{\inf}$ as the false negative rate, we also compute  the  $F_1$-score defined as  $F_1=\frac{2TPR_{\inf}}{2TPR_{\inf}+FPR_{\inf}+FNR_{\inf}}$. Obviously, the larger $F_1$, the better the corresponding method is.

Denote $\hat \beta$ as an estimate of $\beta$ which is based on the full data (FULL), or based on a reduced dataset after HIM is applied (HIM), or a reduced dataset after MIP is applied (MIP). In this paper, we estimate $\beta$ via the Lasso. The accuracy of the estimation is evaluated by  computing $ERR=\|\hat \beta-\beta\|$ and we compare the accuracy of FULL, HIM and MIP.

  Denote the support of $\beta$ as $\mathrm{supp}(\beta)$   and its complement as $\mathrm{supp}(\beta)^c=\{1\cdots, p\}\setminus \mathrm{supp}(\beta)$. We report the success in identifying the support of $\beta$ by reporting
$$TPR_{vs}=\frac{|\mathrm{supp}(\beta)\cap \mathrm{supp}(\hat \beta)|}{|\mathrm{supp}(\beta)|} \  \ \mbox{and}\ \
FPR_{vs}=\frac{|\mathrm{supp}(\beta)^c\cap \mathrm{supp}(\hat \beta)|}{|\mathrm{supp}(\beta)^c|}.$$

In the following simulations, we set $n_{\mathrm{sub}}=n/2+1$. That is, the random subsets $A_r, r=1, \cdots, m,$ all have cardinaltiy $n/2$.   We repeat each experiment $100$ times and report the means of the quantities defined above.  In implementing MIP, we set the number of random subsets as $m=100$ for Example 2. For Example 1, we take $m=100, 200 $ or $300$ to assess the effect of  $m$.
In Table 2,  because the $FPR_{\inf}$ of HIM can be large,  we decided not  to compute the coefficient estimates based on the reduced data to save space  as long as $FPR_{\inf} >0.7$.
Finally, the FDR level is fixed at $\alpha=0.05$.

\subsection{Simulation setup}
      We simulate the data such that there exists a  strong masking effect  in Example 1 and a strong swamping effect in Example 2.    Denote $\mathbf{0}_{s}$ as a $s$-dimensional zero vector  and $\mathbf{1}_{s}$ as a $s$-dimensional vector of $1$'s.

\noindent
{\bf Example 1} (Strong masking effect).  We first generate $n=100$ non-influential observations  from (\ref{Model1}) with $\beta=(0.4, 0.5, 0.5$, $0.6,0.4, \mathbf{0}_{p-5})^\top$.   Let $i_0=\arg\max_{1\le i\le n} |Y_i|$.
We then replace the first $n_{\inf}=10$ non-influential observations by
$$X_{ij}^{\inf}=X_{i_0j}+I(j\in S_i)\cdot i/p, \ \ \  Y_i^{\inf}=Y_{i_0}+\mu+\varepsilon_i^{\inf}\cdot i/p, \ \ \ 1\le j\le p, 1\le i\le n_{\inf}, $$
where $\{S_i\}$, with $|S_i|=10$, are subsets of $\{1,\cdots, 1000\}$ chosen independently with replacement,    and $\varepsilon_i^{\inf}\sim N(0, 0.5)$.
This example is designed such that the influential observations are clustered together and  consequently  many influential observations are  masked by other influential ones.  HIM based on leave-one-out will likely fail to identify many influential points.
 The simulation results are presented in Table \ref{Table1} and plot (a) of Figure \ref{Fig2}.

\begin{table}[!hbp]
\begin{centering}
\caption{\label{Table1}Simulation results of Example 1 with different $\mu$. }
\begin{tabular}{llccccccc}
\hline
     & $\mu$            & 4.0   &   4.5   & 5.0 & 5.5 & 6.0 & 6.5 & 7.0\\\hline
        &  $TPR_{\inf}$  & 0.780 &	0.820 &	0.940 &	0.960 &	1.000 & 1.000 & 1.000 \\
        &  $FPR_{\inf}$  & 0.003 & 	0.005 &	0.004& 	0.003 &	0.003 & 0.002 &	0.002 \\
         &  $F_{1}$     & 0.875 &	0.898 &	0.967 &	0.978 &	0.998 & 0.999 & 0.999 \\
  MIP  &   $ERR$       &0.570 &	0.568 &	0.553 &	0.518 &	0.525 &	0.502 &	0.507\\
$m=100$ %&              &0.169 &	0.147 &	0.158 &	0.155 &	0.126\\
        &  $TPR_{vs}$    & 0.944 &	0.944 &	0.944 &	0.964 &	0.936 &	0.972 &	0.960\\
        &  $FPR_{vs}$    & 0.022 &	0.024 &	0.018 &	0.019 &	0.012 &	0.016 &	0.016\\\cline{2-9}

        &  $TPR_{\inf}$  &0.840 &	0.860 & 0.960    & 0.980 & 1.000 & 1.000 & 1.000 \\
        &  $FPR_{\inf}$  &0.004 &   0.004 &	0.003    & 0.003 &	0.005 & 0.002 &	0.002 \\
        &  $F_1$        & 0.911 &	0.923 &	0.978 &	0.988 &	0.997 & 0.999 & 0.999 \\
MIP    &   $ERR$      &   0.554 &	0.577 & 0.538    & 0.498 &	0.504 & 0.516 &	0.488\\
$m=200$ %&             & 0.179 &	0.150 &	0.144 	 & 0.148 &  0.139\\
        &  $TPR_{vs}$   &  0.964 &	0.948 &	0.972    & 0.972 &  0.980  & 0.948 &  0.960\\
        &  $FPR_{vs}$   &  0.021 &	0.020 & 0.015    & 0.012 &	0.019  & 0.015 &	 0.012\\\hline

       &  $TPR_{\inf}$  &  0.860 &	0.920 &	0.960& 	0.980 &	1.000 & 	1.000 &	1.000\\
       &  $FPR_{\inf}$  &  0.003 &	0.004 &	0.003 &	0.003& 	0.002 &	0.003& 	0.006\\
       &  $F_1 $       & 0.923 &	0.957 &	0.978 &	0.988 &	0.998 & 0.998 & 0.997 \\
  MIP &   $ERR$        & 0.587 &	0.529& 	0.529 &	0.540 &	0.523  &	0.479 &	0.488\\
$m=300$%&               & 0.149 &	0.139 &	0.159 &	0.133& 	0.130 \\
       &  $TPR_{vs}$    &  0.956 &	0.976 &	0.964 &	0.956  &0.956  &	0.972  &0.976\\
       &  $FPR_{vs}$    &  0.024 &	0.019 &	0.019 &	0.016& 	0.015 &	0.016& 	0.013\\\cline{2-9}

     & $TPR_{\inf}$     &  0.040 &	0.280      &0.260   &0.220 &0.460 &0.420 &0.500 \\
     & $FPR_{\inf}$     &  0.067 &   0.048      &0.080 	&0.111 &0.151 	&0.151 &0.147 \\
     &  $F_1$          & 0.072 &	0.421 &	0.388 &	0.331 &	0.571 & 0.535 & 0.607 \\
  HIM& $ERR$           &  0.802 &	0.757      &0.783   &0.848 &0.866  &0.835 &0.856\\
     %&                 & 0.151 &   0.124      &0.171   &0.173 &0.172 \\
     & $TPR_{vs}$       &  0.856 &	0.900      &0.908 	&0.868 &0.816	&0.832 &0.832 \\
     & $FPR_{vs}$       &  0.040 &	0.040      &0.043   &0.044 &0.033  &0.038 &0.036\\\hline

        &  $ERR$      &   0.769 &  0.788 	&0.836 &	0.832 &	 0.885 &	0.895 &	 0.930 \\
  FULL  %&             &  0.136 &  0.125& 	 0.111 &	0.121& 	 0.113\\
        &  $TPR_{vs}$  &   0.948   &0.932 	&0.920 &	0.924 	&0.928 &	0.892 	&0.932\\
        &  $FPR_{vs}$  &   0.047   &0.051 &	 0.055 &	0.052 	&0.055 &0.056 &0.061\\\hline

\end{tabular}

\end{centering}
\end{table}

\noindent
{\bf Example 2} (Strong swamping effect).  We set $\beta=(0.2,0.4,0.5,0.3,0.2,\mathbf{0}_{p-5})^\top$ and
generate  influential observations according to the following scheme. Let $ \mathbf{w}=(w_1,\cdots, w_{20})^\top\in \mR^{20}$ with   $w_j=j\cdot 0.005\mu$. For $i=1,\cdots, n_{\inf}$, we let
\beqr
 &&Y_i^{\inf}=\mathrm{sign}(\sigma_i)\cdot(\tilde\beta^{\top}X_i^{\inf}+\varepsilon_i^{\inf}), \nonumber\\
&& X_i^{\inf}\sim N(\nu_{\inf}, I_p), \ \mbox{with} \ \nu_{\inf}=(\mathbf{0}_{900}^\top, 0.5\mu\mathbf{1}_{100}^\top)^\top\nonumber,\\
&&\tilde\beta=\beta+(\mathbf{0}_{p-20}^\top,\mathbf{w}^\top)^\top,   \nonumber
\eeqr
where $\varepsilon_i^{\inf}\sim N(0,0.5)$  and $\sigma_i$ is a binary variable  with $P(\sigma_i=1)=P(\sigma_i=-1)=1/2$ independent of $(X_i^{\inf}, \varepsilon_i^{\inf})$.
For this example, when $\mu$ is large, there exists a strong swamping effect.  The simulation results are presented in Table \ref{Table2} and plot (b) of Figure \ref{Fig2}.

\begin{table}[!hbp]
\begin{center}
\caption{\label{Table2} Simulation results of  Example 2 with different $\mu$}
\begin{tabular}{llccccccc}
  \hline
   &  $\mu$ & 4 & 5 & 6 & 7 & 8 & 9 & 10\\\hline
%\multicolumn{6}{c}{HIM}\\

     &  $TPR_{\inf}$ & 1.000 & 1.000      & 1.000      & 1.000      & 1.000 & 1.000      & 1.000\\
     &  $FPR_{\inf}$ & 0.003 & 0.004 & 0.007 & 0.014 & 0.000 & 0.000 & 0.000\\
     &  $F_1$       & 0.998 & 0.998      & 0.996      & 0.993      & 0.999 & 1.000      & 0.999\\
MIP&   $ERR$       & 0.253 & 0.252 & 0.264 & 0.256 & 0.269 & 0.252 & 0.248\\
     %&             & 0.137 & 0.057 & 0.059 & 0.056 & 0.049\\
     &  $TPR_{vs}$   & 0.968 & 0.972 & 0.960 & 0.956 & 0.972 & 0.972 & 0.960\\
     &  $FPR_{vs}$   & 0.014 & 0.015 & 0.016 & 0.012 & 0.020 & 0.016 & 0.014\\  \hline

   & $TPR_{\inf}$    & 1.000 &   1.000    & 1.000      & 1.000      & 1.000 & 1.000    & 1.000\\
   & $FPR_{\inf}$    & 0.018 & 0.093 & 0.357 & 0.502 & 0.859 & 0.989 & 1.000\\
    &  $F_1$        & 0.991 & 0.955     & 0.848     & 0.799      & 0.699 & 0.669      & 0.667\\
HIM& $ERR$          & 0.263 & 0.305 & 0.442 & 0.490 & -- & -- & -- \\ %\cline{2-6}
   %&               & 0.095 & 0.100 & 0.189 & 0.176 & 0.132\\\
   & $TPR_{vs}$      & 0.968 & 0.924 & 0.696 & 0.684 & -- & -- & --\\
   & $FPR_{vs}$      & 0.015 & 0.017 & 0.014 & 0.015 & -- & -- & --\\  \hline

      &  $ERR$      & 0.738 & 0.884 & 0.914 & 1.011 & 1.072 & 1.162 & 1.383\\
FULL %&             & 0.147 & 0.197 & 0.463 & 0.381 & 1.114\\
      &  $TPR_{vs}$  & 0.188 & 0.072 & 0.032 & 0.016 & 0.000 & 0.004 & 0.004\\
      &  $FPR_{vs}$  & 0.005 & 0.006 & 0.004 & 0.005 & 0.003 & 0.004 & 0.003\\\hline \end{tabular}
\end{center}
\end{table}

\noindent

\subsection{Summary of the simulation results}
From Table \ref{Table1}--\ref{Table2} and Figure 2, we observe the following phenomena.

 (1). The comparison between HIM and MIP  when  there exists a masking effect (Example 1) or a swamping effect (Example 2).   From Table \ref{Table1} and \ref{Table2} and Figure \ref{Fig2}, we see that  HIM suffers from these effects seriously.  For Example 1, we see that the $TPR_{\inf}$ of HIM is much smaller than that of MIP.  Although its $TPR_{\inf}$ increases as $\mu$ increases,   the increment is slow and  its  $FPR_{\inf}$ increases at the same  time.    For Example 2, we see from Figure \ref{Fig2} that HIM works well when $\mu\in [2,4]$, but HIM suffers from the swamping effect when $\mu$ is large, with its false positive rates much larger than $0.05$.

  On the other hand, MIP performs very well in Example 1 and 2. It is more powerful than HIM with larger $TPR_{\inf}$, while  its $FPR_{\inf}$  is well controlled at the FDR level $\alpha=0.05$.    The price we pay is the computation cost, as $m$ subsets are evaluated in  MIP.  Our simulation  shows that the computing time of MIP increases linearly with   $m$. Therefore choosing a small or moderate $m$ can reduce the computing cost.
   Alternatively,  by noting  that subsets $A_1,\cdots, A_m$ are sampled independently,   the computational time can be reduced if a parallel computing algorithm is used. 
   
(2). From the  comparison between the fit after MIP is used to remove influential points and the fit using the full data, it is clear that  MIP is much better whenever there exist influential observations.  In terms of variable selection, we see that the MIP based fits are slightly better  than the HIM based fits and the FULL data based fits in Example 1. And in Example 2,  the MIP based fits are much better.
Now let us look at the effect of  $m$. From Table \ref{Table1}, we see that MIP performs similarly  for different values of $m$. Using $m=300$ does not bring significant gain over $m=100$.  This shows that MIP may be insensitive to the choice of the number of the subsets.

(3).
{Finally, we compare MIP to the $\Theta$-IPOD method in \cite{She:Owen:2011}. The
 simulation results using  the latter for Example 1 and 2  are summarized in Table \ref{Table8}.  Comparing Table \ref{Table1}--\ref{Table2} with  Table \ref{Table8} leads to the following conclusions. For Example 1, the true positive rates for identifying influential points  are  similar,  but the false positive rates of $\Theta$-IPOD are  much larger than those of MIP.   For Example 2, the $TPR_{\inf}$'s of the $\Theta$-IPOD method are much smaller than those of MIP for every setting, while its $FPR_{\inf}$'s are much larger than  MIP's.  We conclude that MIP is more effective than $\Theta$-IPOD.     Part of the reason may be that the $\Theta$-IPOD method was developed based on a mean shift model,  while  our method does not assume the scheme of influentialness.
}

\begin{table}[!hbp]
\begin{center}
\caption{\label{Table8} Simulation results using the  $\Theta$-IPOD method in \cite{She:Owen:2011}}
\begin{tabular}{llccccccc}
  \hline
            &  $\mu$ & 4.0 & 4.5 & 5.0 & 5.5 & 6.0 & 6.5 & 7.0\\\cline{2-9}
Example 1   &  $TPR_{\inf}$&   0.900&	0.936&	0.980&	0.980&	0.960&	1.000&	1.000      \\
            &  $FPR_{\inf}$&   0.114&	0.093&	0.135&	0.104&	0.125&	0.110&	0.155     \\
            &  $F_1$       &   0.893&   0.922&  0.926&  0.940&  0.920&  0.947&   0.928\\\hline

            &  $\mu$ & 4 & 5 & 6 & 7 & 8 & 9 & 10\\\cline{2-9}
Example 2   &  $TPR_{\inf}$ &0.018&	0.062&	0.092  &  0.116&	0.232&	0.258&	0.382  \\
            &  $FPR_{\inf}$ &0.012&	0.028&	0.035  &  0.032&	0.081&	0.135&	0.203 \\
            & $F_1$         &0.034& 0.113&  0.163 &  0.202&    0.353&  0.370&  0.482\\\hline
\end{tabular}
\end{center}
\end{table}

%==========
\subsection{Real data analysis}

As an illustration, we apply  MIP to detect  influential points in the microarray data from \cite{Chang:etal:2006} which was previously analyzed by Zhao et al. (2013).
  For this dataset, we focus on 120
twelve-week-old male offspring that were selected for tissue harvesting from the
eyes and for microarray analysis. The dataset contains over 31,042
different probe sets.  Following  \cite{Huang:etal:2006}, we take the   probe  gene TRIM32  as the response. This gene  is interesting as it was
  found to cause Bardet-Biedl syndrome, a genetically heterogeneous
disease of multiple organ systems including the retina \citep{Chang:etal:2006}.  One question of
interest in this data analysis is to find genes whose expressions are correlated
with that of gene TRIM32.  We  followed \cite{Huang:etal:2006} to exclude probes
that were not expressed in the eye or that lacked sufficient variation and select $p=1500$ genes  that are mostly correlated with the probe of TRIM32. Therefore, the analysis has $p = 1500$ predictors and a sample size $n = 120$. Before further analysis,
 all the probes are standardized to have mean zero
and standard deviation one  \citep{Huang:etal:2006}.   Applying Lasso to the full data using the default setting of glmnet function in R, we identify 15 significant variables and the $\ell_2$-norm of the estimated coefficient vector equals $0.097$.

\begin{figure}[!bpt]
\caption{\label{Fig3} Comparison between HIM and MIP.  }
\begin{subfigure}[b]{.5\textwidth}
\centering
\includegraphics[width=8cm,height=7cm]{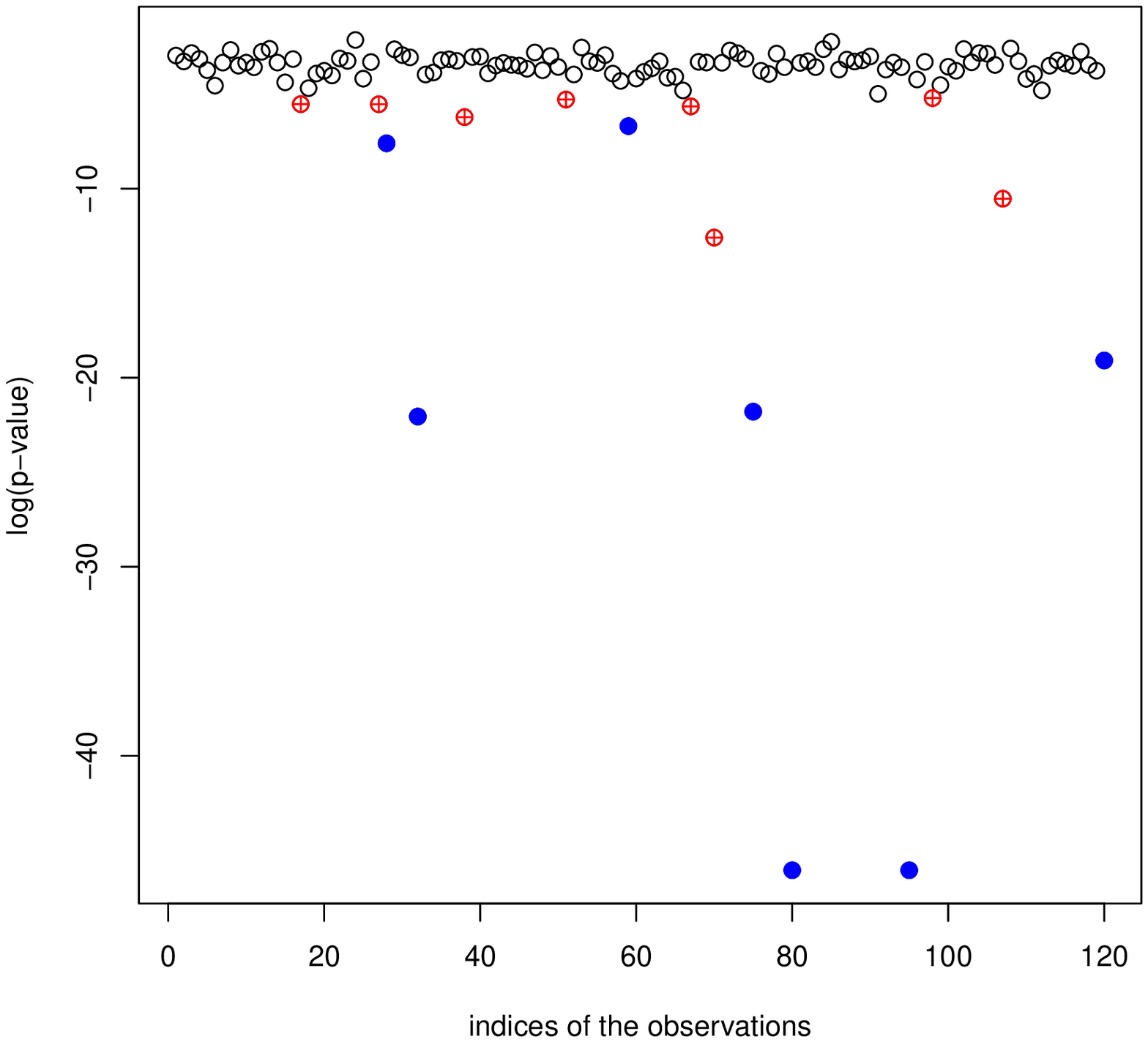}
\caption{The y-axis are the log $p$-values by HIM}
\end{subfigure}
\begin{subfigure}[b]{.5\textwidth}
\centering
\includegraphics[width=8cm,height=7cm]{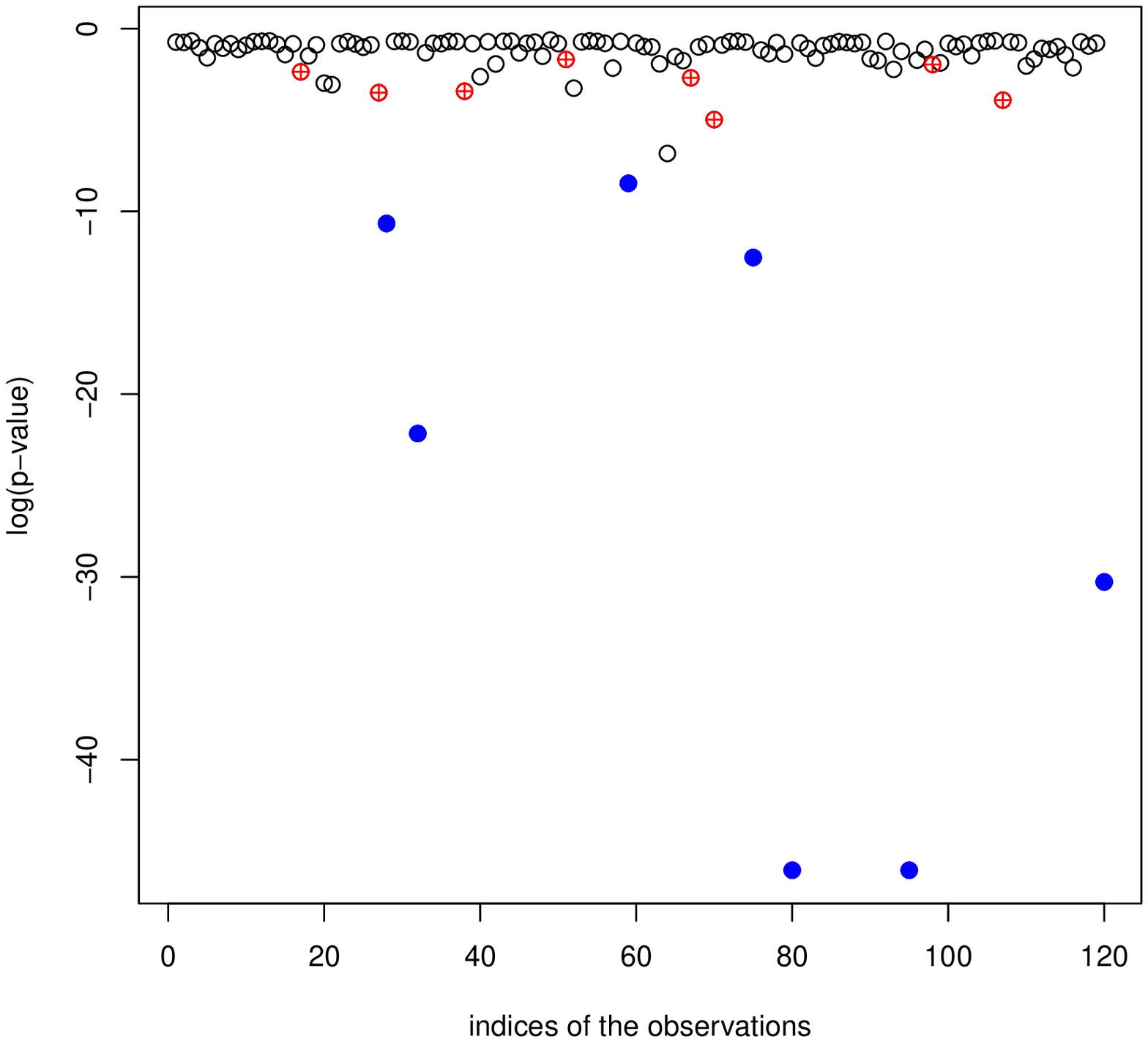}
\caption{The y-axis are the log $p$-values by MIP}
\end{subfigure}
\end{figure}

Applying HIM and MIP to this data with the FDR level at $\alpha=0.05$,  HIM finds 15 influential observations, while  MIP obtains  7 influential observations.
   Interestingly, the set of influential points by MIP is a subset of that by HIM.  In Figure \ref{Fig3},  we plot the influential observations found by MIP  in blue and the extra influential ones by HIM as red crosses, where the y-axis denotes the logarithm of the $p$-values obtained by using HIM as in plot (a) or using MIP as in plot (b).
   Note that, to make the plot more comparable,  the checking step in   the Min-Max-Checking algorithm is applied to  all   observations  such that we can get  a $p$-value for each observation.  From this figure,  we can see that the red crossed points identified by HIM as influential  do not seem to have very small $p$-values.

To make further comparison, we use the ordinary least squares estimation on the important variables found via Lasso, after applying either HIM or MIP, to the non-influential point set identified by HIM. We compare their BIC score defined as BIC=$n\log(RSS/n)+k\log(n)$ where $RSS$ is the residual sum of square, $n=105$ is the same size after removing the $15$ influential points identified by HIM, and $k$ is the number of variables used. Obviously,  a model with   a smaller BIC  is preferred. Note $k=9$ if HIM is used and $k=6$ if MIP is applied. Because of the setup, this comparison favors HIM in some sense. It is found that BIC =$-567.34$ if HIM is applied for influential point detection and BIC=$-578.94$  if MIP is applied. Thus, MIP is potentially more effective for finding a better model than HIM as its BIC value is smaller.

For the real data, of course it is not known which  observations are   influential.
To further assess the performance of HIM and MIP, we artificially add influential points to the dataset and evaluate whether they can find these points afterwards. Specifically,
we first  remove the influential points detected by  each method and  add 10 additional  observations to the remaining  data.  This scheme gives a total of 115 observations for assessing HIM and 123 observations for MIP. The $10$    added influential observations  are generated as
\[X_{iS}=1.1x_S+Z_S, ~X_{iS^c}=x_{S^c}, ~Y_i=1.1y+\epsilon,\ \ \ {1\le i\le 10},\]
where $Z\sim N(0,0.01I_p)$, $S$ is a random subset of $\{1,\cdots, p\}$  consisting of $10$ distinctive indices, $Z_S$ is a  subvector of $Z$ with indices in $S$, $(x,y)$ is chosen  randomly from non-influential point set identified by HIM, and $\epsilon\sim N(0,0.01)$ is independent of $Z$.

 We apply MIP and HIM to the contaminated data defined above  with the nominal FPR set as $0.05$ in the Benjamni-Hochberg procedure
   and repeat the process for $100$ times. Then we  compute the true positive rate (TPR) and false positive rate (FPR) of  the two methods, respectively, for identifying these artificial influential points. It turns out that MIP gives a TPR of $1$ and a FPR of $0.008$, while HIM gives
a TPR of $1$ and a FPR as high as $0.585$. Obviously, HIM suffers  seriously  from the swamping effect  caused by the addition of new influential  observations, while MIP does not seem to be affected by newly added observations.

%=========
\section{Discussion}
We have proposed a novel procedure named MIP for multiple influential point detection in high-dimensional spaces. The MIP procedure is intuitive, theoretically justified, and easy to implement. In particular, by combining the strengths of the Max and Min statistics, the proposed MIP framework can overcome the masking and swamping effects notoriously in influence diagnosis, and is able to identify multiple influential points with prespecified accuracy in terms of false discovery rate control.

Both HIM and MIP are based on the idea of measuring the change in marginal correlations when one observation is removed. The primary consideration for using the marginal correlation is due to its ubiquity in statistical analysis and the possibility of deriving rigorous theoretical results, as we have shown. But it need not be the only quantity that defines  influence. Towards this, it will be interesting to explore using other quantities to define influence for example the generalized OLS estimator used for screening variables in \cite{Wang:Leng:2016}.

Finally, we hope that this paper can bring to the attention of the statistics community the importance of influence diagnosis and how one might think about defining influence and devising automatic procedures for assessing influence, in a theoretically justified fashion. With the rapid advances of the big data analytics, we believe that the issue of influence diagnosis will only become more relevant and hope that this paper can serve as a catalyst to stimulate more research in this area.

%=================

\renewcommand{\baselinestretch}{1}
\normalsize
\bibliographystyle{apalike}

\def\baselinestretch{\spacing}\selectfont

\section*{Appendix}

\subsection*{Proof of Lemma \ref{Lemma1}}
Define $\dot J_r=N_{B_r}^{-2}\|\sum\limits_{t\in B_r}\dot Y_t\dot\bX_t^\top\|^2$. Observe
  $$J_{\max,n}\le \max\limits_{1\le r\le m} |J_r-\dot J_r|+\max\limits_{1\le r\le m} \dot J_r.$$
   The main idea of the proof is to show that the two terms on the righthand are small.
 For simplicity, we assume that each element of $(\bX, Y)$ has   population mean 0 and variance 1, that is,  $\sigma_{xj}=\sigma_{y}=1$ and $\mu_{xj}=\mu_{y}=0, 1\le j\le p$.
 Before the proof, we review some facts. For any $1\le t_1, t_2\le n$, define
  $$\hat K_{p,t_1t_2}=p^{-1}\hat\bX_{t_1}^\top\hat\bX_{t_2},\  K_{p,t_1t_2}=p^{-1}\dot\bX_{t_1}^\top\dot\bX_{t_2},\quad
  \hat F_{t_1t_2}=\hat Y_{t_1}^\top\hat Y_{t_2},\quad   F_{t_1t_2}=\dot Y_{t_1}^\top\dot Y_{t_2}.$$
 Then by Lemma 1 of Zhao et al. (2013), we have $E(K_{p,t_1t_2})=0$ if $t_1\ne t_2$ and $1$ if $t_1=t_2$.
 Besides,  $E(K_{p,tt}-1)^2=O(p^{-2}l_p)$ and $E(K_{p,t_1t_2})^2=O(p^{-2}l_p)$, for any $t_1\ne t_2$.
In addition, $F_{tt}\sim \chi^2(1)$ due to $\dot Y_t\sim N(0,1)$.

{\bf Part  I.}\ \  We show  $\max\limits_{1\le r\le m} |J_r-\dot J_r|=O_p((\log (np))(\log n)(\log p)n^{-1/2})$.

{\it Step 1}. We first simplify  the expression of $\hat \bX_t$ and $\hat Y_t$.

It is easy to see that   for  $1\le t\le n$,
   $$\hat X_{tj}=\dot X_{tj}\frac{\sigma_{xj}}{\hat\sigma_{xj}}+ \frac{\mu_{xj}-\hat\mu_{xj}}{\sigma_{x_j}} \frac{\sigma_{x_j}}{\hat\sigma_{xj}}:=\dot X_{tj}(1+w_{x,nj}^{(1)})+w_{x,nj}^{(2)},$$
where $w_{x,nj}^{(1)}=(\frac{\sigma_{xj}}{\hat\sigma_{xj}}-1)$ and  $w_{x,nj}^{(2)}=\frac{\mu_{xj}-\hat\mu_{xj}}{\sigma_{x_j}}  \frac{\sigma_{x_j}}{\hat\sigma_{xj}}=\frac{\mu_{xj}-\hat\mu_{xj}}{\sigma_{x_j}} + \frac{\mu_{xj}-\hat\mu_{xj}}{\sigma_{x_j}} (\frac{\sigma_{x_j}}{\hat\sigma_{xj}}-1)$.
Let $w_{x,n}^{(1)}=\max\limits_{1\le j\le p} |w_{x,nj}^{(1)}|$ and $w_{x,n}^{(2)}=\max\limits_{1\le j\le p} |w_{x,nj}^{(2)}|$.
 By (C5) and simple calculation, we have, for some constant $0<C<\infty$,
 \beq\label{eq_uni_sigm_mu}
 P(n^{1/2} w_{x,n}^{(1)}>C\log p)\le p^{-3}, \quad p(n^{1/2}  w_{x,n}^{(2)}>C\log p)\le p^{-3}.
 \eeq
 That is
 \beq\label{eq_max_sigm_mu}
 w_{x,n}^{(1)}=O_p((\log p)n^{-1/2}),\ \ \   w_{x,n}^{(2)}=O_p((\log p)n^{-1/2}).
 \eeq
Similarly, let  $\dot Y_t=\sigma_y^{-1}(Y_t-\mu_t)$, which follows standard normal $N(0,1)$. Then
\beqr
\hat
Y_t&=&\hat\sigma_y^{-1}(Y_t-\hat\mu_{y})=\dot Y_t+(\hat\sigma_y^{-1}\sigma_y-1)\dot Y_t+\hat\sigma_y^{-1}(\mu_y-\hat\mu_y)\nonumber\\
&:=&\dot Y_t+U_n\dot Y_t+\mathbf{v}_n,
\eeqr
where $U_n$ and $\mathbf{v}_n$ are defined accordingly.  Let $w_{y}^{(1)}=U_n^2, w_{y}^{(2)}=\|\mathbf{v}_n\|$.
Note that $w_y^{(1)}=O_p(n^{-1})$ according to the assumption on $S_{R_y}$ in  (C5).  Similarly, we have $w_{y}^{(2)}=O_p(n^{-1/2})$ by (C5).

{\it Step 2.}\ \ Simplify the expression of $\max_{1\le r\le m}|J_r-\dot J_r|$.

 Recall the definition of $\hat K_{p,t_1t_2}$, $K_{p,t_1t_2}$, $\hat F_{p,t_1t_2}$ and  $ F_{p,t_1t_2}$. Define
$$A_{t_1t_2}=\hat F_{p,t_1t_2}\hat K_{p,t_1t_2}-F_{p,t_1t_2}K_{p,t_1t_2}, \quad  1\le t_1, t_2\le n. $$
 The we have
\beqr
|A_{t_1t_2}| &\le&  |\hat K_{p,t_1t_2}| | \hat F_{p,t_1t_2}-F_{p,t_1t_2}|+
 | F_{p,t_1t_2}| |\hat K_{p,t_1t_2}- K_{p,t_1t_2}|.\nonumber
\eeqr
By Assumption (C1), we see that $N_{B_r}>\delta_1 n$ for all $1\le r\le m$, that is,  $N_{B_r}$ has the same order as $n$.
By simple calculations, we have
$$J_r-\dot J_r=N_{B_r}^{-2}\left\{\sum\limits_{t\in B_r}A_{tt}+\sum\limits_{t_1\ne t_2,t_1,t_2\in B_r}^nA_{t_1t_2}\right\}.$$
Then it follows that
\beqr\label{max_{R-dotR}}
\max_{1\le r\le m} |J_r-\dot J_r|&=&\max_{r}N_{B_r}^{-2}\left|\sum\limits_{t\in B_r}A_{tt}+\sum\limits_{t_1\ne t_2,t_1,t_2\in B_r}^nA_{t_1t_2}\right|\nonumber\\
&\le&\max_{r}\left\{N_{B_r}^{-1}\max\limits_{1\le t\le n}|A_{tt}|+\frac{N_{B_r}-1}{N_{B_r}}\max\limits_{t_1\ne t_2,1\le t_1,t_2\le n}A_{t_1t_2}\right\}\nonumber\\
&\le&\max_{r}\left\{ N_{B_r}^{-1}\max\limits_{1\le t\le n}|A_{tt}|+\max\limits_{t_1\ne t_2,1\le t_1,t_2\le n}|A_{t_1t_2}|\right\}.\nonumber\\
&\le&(\max_r N_{B_r}^{-1})\max\limits_{1\le t\le n}|A_{tt}|+\max\limits_{t_1\ne t_2,1\le t_1,t_2\le n}|A_{t_1t_2}|.\nonumber\\
% &=&  O_p([\log(n)\log (p)]^{1/2}n^{-1/4})+O_p(n^{-3/2}\log n)+O_p(p^{-1}l_p^{1/2}\log n) \nonumber
\eeqr

{\it Step 3}. We study the terms in $A_{t_1t_2}$.

{\it Step 3.1}. We show $\max_{t_1t_2}F_{t_1t_2}=O_p(\log n)$ and
$$\max_{t_1,t_2}|\hat F_{p,t_1t_2}-F_{p,t_1t_2}|=O_p((\log n)/\sqrt{n}).$$

 Because $\dot Y_t$'s   are {\it i.i.d.} variables with distribution $N(0,1)$,
 $\|\dot Y_t\|^2\sim \chi^2(1)$ and consequently, by the tail probability of  $\chi^2(1)$ distribution, we have
$\max_t\|\dot Y_t\|^2=O_p(\log n)$. By Cauchy-Schwarz inequality, we see that  $\max_{t_1t_2}F_{t_1t_2}=O_p(\log n)$ holds.
 In addition, by the results in Step 1, applying  Cauchy-Schwarz inequality and triangle inequality, we have
\beqr
\max_t\|\hat Y_t-\dot Y_t\|^2&\le& \max_t\|U_n\dot Y_t+\mathbf{v}_n\|^2\le 2[ (\max_t\|\dot Y_t\|^2)w_{y}^{(1)}+(w_{y}^{(2)})^2]\nonumber\\
&=&O_p((\log n)/n) +O_p(n^{-1})=O_p((\log n)/n).
\eeqr
Moreover, for any $1\le t_1,t_2\le n$,
\beqr
\max_{t_1,t_2}|\hat F_{p,t_1t_2}-F_{p,t_1t_2}|&=&\max_t|\dot Y_{t_1}^\top(\hat Y_{t_2}-\dot Y_{t_2})+(\hat Y_{t_1}-\dot Y_{t_1})^\top\dot Y_{t_2}+(\hat Y_{t_1}-\dot Y_{t_1})^\top(\hat Y_{t_2}-\dot Y_{t_2})|\nonumber\\
&\le & 2\max_{t_1t_2}|\dot Y_{t_1}^\top(\hat Y_{t_2}-\dot Y_{t_2})|+\max_{t_1t_2}|(\hat Y_{t_1}-\dot Y_{t_1})^\top(\hat Y_{t_2}-\dot Y_{t_2})|\nonumber\\
&\le & 2\max_{t_1}\|\dot Y_{t_1}\| \max_{t_2}\|\hat Y_{t_2}-\dot Y_{t_2}\|+\max_{t_1}\|\hat Y_{t_1}-\dot Y_{t_1})\|^2 \nonumber\\
&=&O_p((\log n)/\sqrt{n}).\eeqr

{\it Step 3.2}. We show
$$\max\limits_{t_1,t_2}|\hat K_{p,t_1t_2}-K_{p,t_1t_2}|=O_p(\log (pn)(\log p)/\sqrt{n}).$$
In fact, it is easy to see
 \beqr
\max_{t_1,t_2}|\hat K_{p,t_1t_2}-K_{p,t_1t_2}|&=&\max_{t_1,t_2}|p^{-1}[\bX_{t_1}^\top(\hat \bX_{t_2}-\bX_{t_2})+(\hat \bX_{t_1}-\bX_{t_1})^\top \bX_{t_2}+(\hat \bX_{t_1}-\bX_{t_1})^\top(\hat \bX_{t_2}-\bX_{t_2})]|.\nonumber\\
\eeqr
For any  $1\le t_1,  t_2\le n$, we have
$$p^{-1}\max_{ 1\le t\le n }|\bX_{t_1}^\top(\hat \bX_{t_2}-\bX_{t_2})|\le \max_{1\le j\le p,1\le t\le n } |X_{t_1j}X_{t_2j}|w_{x,n}^{(1)}+  \max_{1\le j\le p,1\le t\le n }|X_{tj}| w_{x,n}^{(2)}. $$
Since $X_{tj}$ are standard normal and $X_{tj}$'s  are independent with respect to $1\le t\le n$, we have
$$ \max_{1\le j\le p,1\le t\le n }|X_{tj}| =O_p((\log(pn))^{1/2}),$$
$$\max_{1\le j\le p,1\le t\le n }  |X_{t_1j}X_{t_2j}|\le \max_{1\le j\le p,1\le t\le n }  |X_{t_1j}|\max_{1\le j\le p,1\le t\le n }  |X_{t_2j}|=O_p(\log (pn)).$$
Combining with  (\ref{eq_max_sigm_mu}) in  Step 1, we have $p^{-1}\max_{t_1,t_2}|\bX_{t_1}^\top(\hat \bX_{t_2}-\bX_{t_2})|=O_p(\log (pn)(\log p)/\sqrt{n})$.
By similar arguments and noting $(\log p)/\sqrt{n}=o(1)$, we have
\beqr
p^{-1}|(\hat \bX_{t_1}-\dot\bX_{t_1})^\top(\hat \bX_{t_2}-\dot\bX_{t_2})|&\le &\max\limits_{1\le j\le p} |(\hat X_{t_1j}-\dot X_{t_1j})(\hat X_{t_2j}-\dot X_{t_2j})| \nonumber\\
&\le &  \left[\max\limits_{1\le j\le p}|\dot X_{t_1j}| w_{x,n}^{(1)}+w_{x,n}^{(2)}\right]\left[\max\limits_{1\le j\le p}|\dot X_{t_2j}| w_{x,n}^{(1)}+w_{x,n}^{(2)}\right]\nonumber\\
&=&O_p(\log (pn)(\log p)/\sqrt{n}).
\eeqr
Therefore, we prove the conclusion on $\max_{t_1,t_2}|\hat K_{p,t_1t_2}-K_{p,t_1t_2}|$.

{\it Step 3.3}. We show
$\max\limits_{t_1,t_2}|\hat K_{p,t_1t_2}|=O_p(\log (np))$.

Note
$$\max\limits_{t_1,t_2}|\hat K_{p,t_1t_2}|\le  \max\limits_{t_1,t_2}|K_{p,t_1t_2}|+ \max\limits_{t_1,t_2}|\hat K_{p,t_1t_2}-K_{p,t_1t_2}|.$$
The second term has been analyzed in Step 3.2. Consider the first term which satisfies
$$\max\limits_{t_1,t_2}|K_{p,t_1t_2}|\le \max\limits_{t_1,t_2}|E(K_{p,t_1t_2})|+\max\limits_{t_1,t_2}|K_{p,t_1t_2}-E(K_{p,t_1t_2})|.$$
Since $X_{tj}$'s are standard normal, we have  arguments similar to before that
$$\max\limits_{t_1,t_2}|K_{p,t_1t_2}-E(K_{p,t_1t_2})|\le \max\limits_{t_1,t_2,j}|\dot X_{tj}\dot X_{t_2j}-E(\dot X_{tj}\dot X_{t_2j})|=O_p(\log(np)).$$
For $E(K_{p,t_1t_2})$,  recall that $E(K_{p,tt})=1$ and $E(K_{p,t_1t_2})=0$ if $t_1\ne t_2$.
Thus,  we have the conclusion of Step 3.3. Finally, combining all the results in Step 3, it follows that
$$\max_{t_1,t_2}A_{t_1t_2}=O_p(\log(np)\log(n) n^{-1/2}+\log(np)(\log n)(\log p)n^{-1/2})=O_p((\log (np))(\log n)(\log p)n^{-1/2}).$$
Combining with (\ref{max_{R-dotR}}), we have the conclusion of Step 3 and it follows that
$$\max_{1\le r\le m} |J_r-\dot J_r|=O_p((\log (np))(\log n)(\log p)n^{-1/2}).$$
This complets the proof of Part I.

{\bf Part II}.  We show the final conclusion  by considering  $\max_{1\le r\le m} \dot J_r$. Note
\beqr
 \dot J_r=N_{B_r}^{-2}[\sum\limits_{t_1\in B_r} F_{t_1t_1} K_{p,t_1t_1} +\sum\limits_{t_1,t_2\in B_r, t_1\ne t_2} F_{t_1t_2} K_{p,t_1t_2} ].\nonumber
 \eeqr
Then
\beqr
\max_r  |\dot J_r|&=&\max_{r}N_{B_r}^{-2}[\sum\limits_{t_1\in B_r} |F_{t_1t_1} K_{p,t_1t_1}| +\sum\limits_{t_1,t_2\in B_r, t_1\ne t_2} |F_{t_1t_2} K_{p,t_1t_2}| ]\nonumber\\
&\le &[\min_r N_{B_r}]^{-2}\left[\sum\limits_{1\le t_1\le n} |F_{t_1t_1} K_{p,t_1t_1}| +\sum\limits_{1\le t_1,t_2\le n, t_1\ne t_2} |F_{t_1t_2} K_{p,t_1t_2}| \right].\nonumber
 \eeqr
Note $N_{B_r}>\delta_1n$.
Then by Cauchy-Schwarz inequality, we have
\beqr
T_1&:=&E\{[\min_r N_{B_r}]^{-2}\sum\limits_{1\le t_1\le n} |F_{t_1t_1} K_{p,t_1t_1}|\}\nonumber\\
&\le& (n\delta_1)^{-2} n E|F_{t_1t_1} K_{p,t_1t_1}|\nonumber\\
&\le& (n\delta_1)^{-2} n [E(F_{t_1t_1})^2]^{1/2} [E(K_{p,t_1t_1})^2]^{1/2}.
\eeqr
Noting that $F_{tt}\sim \chi^2(1)$, we have that $E(F_{t_1t_1})^2$ is bounded.  Moreover, noting $E(K_{p,tt})=1$ and $E(K_{p,tt}-1)^2=O(p^{-2}l_p)$, we have
$$E(K_{p,t_1t_1})^2= E(1+K_{p,t_1t_1}-1)^2\le 1+E(K_{p,tt}-1)^2 =1+O_p(p^{-2}l_p).$$
Then $T_1=O(n^{-1})$.
Similarly, we have
\beqr
T_2&:=&E\left\{[\min_r N_{B_r}]^{-2}\sum\limits_{1\le t_1,t_2\le n, t_1\ne t_2} |F_{t_1t_2} K_{p,t_1t_2}|\right\}\nonumber\\
&\le& (n\delta_1)^{-2}n(n-1) E|F_{t_1t_2} K_{p,t_1t_2}|\nonumber\\
&\le& \delta_1^2 [E(F_{t_1t_2})^2]^{1/2} [E(K_{p,t_1t_2})^2]^{1/2},
\eeqr
where $t_1\ne t_2$ in the second inequality.
By the Cauchy-Schwarz inequality, we have  $E(F_{t_1t_2})^2\le [E(F_{tt})]^2<\infty$. On the other hand,
$E(K_{p,t_1t_2})^2=O(p^{-2}l_p)$. Therefore, $T_2=O(p^{-1}l_p^{1/2})$.
Combining, we have $\max\limits_{1\le r\le m}  |\dot J_r|=O_p(p^{-1}l_p^{1/2})$. Finally combining the conclusions in Part I and Part II, we have
$$J_{\max,n}=\max_{1\le r\le m} J_r=O_p\left(\log(np)(\log n)(\log p)n^{-1/2}+p^{-1}l_p^{1/2}\right).$$

\subsection*{Proof of Theorem \ref{Th2}}
Recall $n_{\mathrm{sub}}=k_{\mathrm{sub}}n$.     Simple calculations shows that
$$\mD_{r,k}=p^{-1}\|\hat\brho_{A_r^{(+k)}}-\hat\brho_{A_r}\|^2
 =p^{-1}\|\frac{1}{n_{\mathrm{sub}}(n_{\mathrm{sub}}-1)}\sum\limits_{t\ne k,t\in A_r }  \hat  Y_t\hat \bX_{t}^\top-\frac{1}{n_{\mathrm{sub}}}\hat Y_k\hat \bX_{k}^\top \|^2.\nonumber
$$
Consequently, it holds that
 \beqr
n_{\mathrm{sub}}^2\mD_{r,k}&=&p^{-1}\|\frac{1}{n_{\mathrm{sub}}-1}\sum\limits_{t\in B_r\setminus\{k\} }  \hat Y_t\hat \bX_{t}^\top- \hat Y_k\hat \bX_{k}^\top \|^2\nonumber\\
&:=&p^{-1}\|W_{r,non}-\hat Y_k\hat \bX_{k}^\top  \|^2.\nonumber
 \eeqr
By Lemma \ref{Lemma1}, we have $p^{-1}\max\limits_{1\le r\le m }\|W_{r,non}\|^2=O_p(\xi_{n,p}+p^{-1}l_p^{1/2})$.
Therefore,
\beqr
\max_{1\le r\le m}n_{\mathrm{sub}}^2\mD_{r,k}&=&p^{-1}\|\hat Y_k\hat \bX_{k}^\top  \|^2(1+   O_p(\xi_{n,p}+p^{-1}l_p^{1/2})).\nonumber
%&\le &2 p^{-1}\left\{\|\dot Y_k\dot\bX_k^\top\|^2+\|\hat Y_k\hat \bX_{k}^\top-\dot Y_k\dot\bX_k^\top  \|^2\right\}(1+   O_p(\xi_{n,p}+p^{-1}l_p^{1/2})).\nonumber
\eeqr
On the other hand, based on assumption (C5) and the proof of Lemma \ref{Lemma1}, we have
$p^{-1}\|\hat Y_k\hat \bX_{k}^\top -\dot Y_k\dot \bX_{k}^\top \|^2\le \max_{j,s}|\hat Y_{ks}\hat X_{kj} -\dot Y_{ks}\dot X_{kj}|^2=o_p(1)$.
That is,  $p^{-1}\|\hat Y_k\hat \bX_{k}^\top\|^2=p^{-1}\|\dot Y_k\dot \bX_{k}^\top\|^2(1+o_p(1))$.
Furthermore, note  that $p^{-1}\|\dot \bX_{k}\|^2=K_{p,tt}$ and that  $E(K_{p,tt}-1)^2=O(p^{-2}l_p)$. It follows that  $p^{-1}\|\dot \bX_{k}\|^2=O_p(1)$.
Consequently, we have
$$p^{-1}\|\dot Y_{k}\dot \bX_{k}\|^2=\|\dot Y_{k}^2\|^2  \left(p^{-1} \|\dot \bX_{k}\|^2\right)=\|\dot Y_{k}\|^2(1+o_p(1)). $$
Note that  $\dot Y_{k}$ follows $N(0,1)$. Therefore,
$$T_{\max,k}=\max_{1\le r\le m}n_{\mathrm{sub}}^2\mD_{r,k}=\|\dot Y_{k}\|^2(1+o_p(1))+o_p(1).$$
Consequently, $T_{\max,k}\rightarrow_d \chi^2(1)$.
By nearly the same argument, it is easy to see that  $T_{\min,k}\rightarrow_d \chi^2(1)$.
This completes the proof. $\blacksquare$

\subsection*{Proof of Theorem \ref{Th3}}

(1) We first prove  the conclusion that  $F_{\max,k}\le R_{\inf}^2 d_{S_{\inf}}$ mentioned just before Theorem \ref{Th3}.
    Note that  $n_{\inf}=n\delta_{\inf,n}$ and that $R_{\inf}=\delta_{\inf,n}/k_{\mathrm{sub}}$ where  $k_{\mathrm{sub}}>0$ by assumption  (C1). Denote  $\tilde W_{\inf,k,r}=n_{\inf}^{-1}\sum\limits_{t\in O_r} \hat Y_t\hat \bX_t^\top$. Obviously we have   $0\le |O_r|\le n_{\inf} $, due to  the fact $O_r\subseteq S_{\inf}\setminus\{k\}$.  Recall the definition of   $d_{S}$.  Then
$$
p^{-1}\max\limits_{1\le r\le m}\|\tilde W_{\inf,k,r}\|^2 \le \max_{1\le r\le m}\max\limits_{t\in O_r}E_t\le \max\limits_{t\in S_{\inf}\setminus\{k\}}E_t =d_{S_{\inf}\setminus\{k\}}.
$$
Recall that $F_{\max,k}=p^{-1}\max\limits_{1\le r\le m}\|W_{\inf,k,r}\|^2$. Then, it holds  that
\beqr\label{eq3.3}
F_{\max,k}= R_{\inf}^2\cdot p^{-1}\max\limits_{1\le r\le m}\|\tilde W_{\inf,k,r}\|^2\le R_{\inf}^2 d_{S_{\inf}\setminus\{k\}}\le R_{\inf}^2 d_{S_{\inf}}.
\eeqr

(2) We prove the conclusion of ($i$) and ($ii$).
  Recall   that   $n_{\mathrm{sub}}^2\mD_{r,k}=p^{-1}\|W_{non,k,r}+W_{\inf,k,r}-\hat Y_k\hat \bX_{k}^\top  \|^2$ by (\ref{dec_D_rk}).  By Lemma \ref{Lemma1}, it follows  that    $J_{\max,n}=p^{-1}\max\limits_{1\le r\le m}\|W_{non,k,r}\|^2=O_p(\xi_{n,p}+p^{-1}l_p^{1/2})=o_p(1)$.
Consequently, by the Cauchy-Schwarz inequality, it holds that
 \beqr\label{App_Tmin}
T_{\min,k} &=&\min_{1\le r\le m} n_{\mathrm{sub}}^2\mD_{r,k}\nonumber\\
 &=&\min_{1\le r\le m} p^{-1}\|W_{\inf,k,r}-\hat Y_k\hat \bX_{k}^\top\|^2(1+o_p(1)),
\eeqr and
 \beqr\label{App_Tmax}
T_{\max,k} &=&\max_{1\le r\le m} n_{\mathrm{sub}}^2\mD_{r,k}\nonumber\\
 &=&\max_{1\le r\le m} p^{-1}\|W_{\inf,k,r}-\hat Y_k\hat \bX_{k}^\top\|^2(1+o_p(1)).
\eeqr

We prove the conclusion in (i).   As $F_{\max,k}\rightarrow 0$, we have
$ T_{\min,k}$ and $ T_{\min,k}$ converge in probability to $p^{-1}\|\hat Y_k\hat \bX_{k}^\top\|^2(1+o_p(1))$.  When $Z_k$ is non-influential, by the proof of Theorem \ref{Th2}, we have $p^{-1}\|\hat Y_k\hat \bX_{k}^\top\|^2=E_k\rightarrow_d \chi^2(1)$.

We prove the conclusion in (ii). Due to the definition of $F_{\min,k}$, we can always find  some  $r_0=r_0(m)$ such that
$F_{\min,k}=p^{-1}\|W_{\inf,k,r_0}\|^2$.  When $Z_k$ is influential, by (\ref{App_Tmax})   and the definition of $E_{k}$,  it follows  that
\beqr
T_{\max,k}^{1/2}&\ge &[n_{\mathrm{sub}}^2\mD_{r_0,k}]^{1/2}=p^{-1/2}(\|\hat Y_k\hat \bX_{k}^\top\|-\|W_{\inf,k,r_0}\|)(1+o_p(1))^{1/2}\nonumber\\
&= &(E_k^{1/2}-F_{\min,k}^{1/2})(1+o_p(1))^{1/2}.\nonumber
\eeqr
 Since    $E_{k}^{1/2}-F_{\min,k}^{1/2}>(\chi^2_{1-\alpha}(1))^{1/2}$, we have $P(T_{\max,k}>\chi^2_{1-\alpha}(1))\rightarrow 1$.
This completes the proof. $\blacksquare$

\subsection*{Proof of Proposition \ref{Proposition2}}
 Note that $J_r$ is defined for fixed $k$, that is, $J_r$ depends on $k$.    Checking  Step 2 of Part I and Part II in the proof of Lemma \ref{Lemma1}, we see that both $\max_{1\le r\le m} |J_r-\dot J_r|$  and $\max_{1\le r\le m} |\dot J_r|$ have upper   bounds independent of $k$. Therefore, Lemma \ref{Lemma1} actually holds uniformly over $k$,  that is,    $\max\limits_{k} J_{\max,n}=O_p(\xi_{n,p}+p^{-1}l_p^{1/2})=o_p(1)$.

 By the proof of (ii) in the proof of Theorem \ref{Th3},
  $T_{\max,k}^{1/2}>(E_k^{1/2}-  F_{\min,k}^{1/2})(1+o_p(1))^{1/2}$, where  the term   $o_p(1)$ depending  on $\max\limits_{k} J_{\max,n}$ is independent of $k$.
 Therefore, $P(\cap_{k\in S_{\inf}} \{T_{\max,k}^{1/2}>E_k^{1/2}-F_{\min,k}^{1/2}\})\rightarrow 1 $.
Note that $\min\limits_{k\in S_{\inf}} T_{\max,k}^{1/2}>\min\limits_{k\in S_{\inf}} E_{k}^{1/2}-\max\limits_{k\in S_{\inf}} F_{\min,k}^{1/2}$.
  Since $a_0$ is independent of $k$, we have $\max\limits_{k\in S_{\inf}} F_{\min,k}<a_0^{2}$.  Consequently, according to the assumption  $E_k^{1/2}>(\chi^2_{1-\alpha}(1))^{1/2}+a_0$,   we have
  $P(\min\limits_{k\in S_{\inf}} T_{\max,k}>\chi^2_{1-\alpha}(1)  )\rightarrow 1.$  Since $\chi^2(1)$ is the limit distribution under the null hypothesis of  no influential observations, the $p$-values  associated with observations of  indices in set $S_{\inf}$ are no more than $\alpha$ in probability. Therefore $\max\limits_{k\in S_{\inf}} p_{\max,k}<\alpha$ with probability tending to 1.

 Recall that $p_{\max,(i)}$'s are the increasing order of $p$-value $p_{\max,i}$'s.  Let $k'$ be the largest $i$ such that $p_{\max,(i)}\le \alpha_0i/n$.  The Benjamini-Hochberg procedure rejects  hypothesis $H_{0(i)}$, where  $1\le i\le k'$.
Denote by $[i]$ as the rank  of $p_{\max,i}$ in the  series $p_{\max,(i)}$'s. Let $\max\limits_{i\in S_{\inf}}[i]$ be the largest rank for $p_{\max,i}, i\in S_{\inf}$.
 If $\max\limits_{i\in S_{\inf}}p_{\max,i}$ is less than $\alpha_0 \max\limits_{i\in S_{\inf}}[i]/n$ for $i\in S_{\inf}$, then according to the rejection rule of the Benjamini-Hochberg procedure, all $H_{0i}$ with $i\in S_{\inf}$ will be rejected.
 Noting   that    $\alpha=\alpha_0 \delta_{\inf,n}$, we have  in probability tending to one
 $$\max\limits_{i\in S_{\inf}} p_{\max,i}\le \alpha=\alpha_0\delta_{\inf,n}=\alpha_0n_{\inf}/n.$$
On the other hand, it is easy to see that  $\max\limits_{i\in S_{\inf}}[i]\ge n_{\inf}$.
Thus, it follows that $\max\limits_{i\in S_{\inf}}p_{\max,i}\le  \max\limits_{i\in S_{\inf}}\alpha_0[i]/n$ in probability tending to 1.
Therefore, all $H_{0i}$ with $i\in S_{\inf}$ will be rejected by the Benjamini-Hochberg procedure. $\blacksquare$

\subsection*{Proof of Theorem \ref{Th4} and Proposition \ref{Prop3}}
The proof of Theorem \ref{Th4} is similar to that of Theorem \ref{Th3}.  We first prove the conclusion in (i) of Theorem \ref{Th4}.   By (\ref{App_Tmin}) and as $F_{\min,k}\rightarrow 0$, we see that $T_{\min,k}\rightarrow_p E_k$ and that $E_k\rightarrow_d\chi^2(1)$ for any $k\in S_{\inf}^c$.
Now we turn to conclusion (ii)  of Theorem \ref{Th4}. Note that $T_{\min,k}^{1/2}>(E_k^{1/2}-  F_{\max,k}^{1/2})(1+O_p(\xi_{n,p}+p^{-1}l_p))^{1/2}=(E_k^{1/2}-  F_{\max,k}^{1/2})(1+o_p(1))^{1/2}$. According to the argument in the proof of Proposition  \ref{Proposition2},  the  term $O_p(\xi_{n,p}+p^{-1}l_p)$  is  independent of $k\in S_{\inf}$.
 Therefore, $P( T_{\min,k}^{1/2}>E_k^{1/2}-F_{\max,k}^{1/2})\rightarrow 1 $.
Combining  with the assumption $E_k^{1/2}> F_{\max,k}^{1/2}+(\chi^2_{1-\alpha}(1))^{1/2}$, we have the conclusion as desired.

Finally, we prove Proposition \ref{Prop3}.   Recall that $F_{\max,k}\le R^{2}_{\inf}d_{S_{\inf}\setminus\{k\}}$ in (\ref{eq3.3}).  The sufficient condition in Proposition \ref{Prop3} is derived from the fact that $d_{S_{\inf}\setminus\{k\}}\le E_{(1)}$ and the Min-Unmask condition of Theorem \ref{Th4}.
$\blacksquare$

\subsection*{Proof of Proposition \ref{Proposition3}}
  We consider only the  case when $K=2$.  The proof of the general case is similar. Denote by $n_1$ and $n_2$ as the {expected} number of hypothesis rejected in round 1 and 2, respectively. Since the FDR level is controlled at $\alpha_0$ in each round, then for estimate  $S_{\min}^1\cup S_{\min}^2$,  the {expected} number of falsely rejected hypotheses is less than $\alpha_0 (n_1+n_2)$ where $n_1+n_2$ is the expectation of the  total number of rejected ones. Therefore FDR is still controlled at level $\alpha_0$, that is, $\tilde R_{non}/(\tilde R_{non}+\tilde R_{inf})\le \alpha_0$, where $\tilde R_{non}$ is the expected  number of non-influential observations that are falsely labeled  as influential ones,  and  $\tilde R_{inf}$ is the expected  number of influential observations that are correctly identified. Due to the fact $\tilde R_{inf}\le n\delta_{\inf,n}$, we have $\tilde R_{non}\le \alpha_0(1-\alpha_0)^{-1}n\delta_{\inf,n}$. Then
  $$E(FPR(\hat\mS))=  \frac{\tilde R_{non}}{n(1-\delta_{\inf,n})} \le \frac{\alpha_0\delta_{\inf,n}}{(1-\alpha_0)(1-\delta_{\inf,n})}\le \frac{\alpha_0}{1-\alpha_0},$$
where  we use in the last equality  the assumption that $\delta_{\inf,n}<1/2$ in (C1).
  $\blacksquare$

\end{document}